\documentclass[aps, prd, superscriptaddress, preprintnumbers, floatfix, longbibliography, nofootinbib, twocolumn, 10pt]{revtex4-2}
\usepackage{epsfig}
\usepackage{enumitem}

\usepackage{graphicx,amsfonts,amsmath,amssymb,amstext,bm,dsfont}
\usepackage{float,wrapfig}
\usepackage{subfigure,psfrag}
\usepackage{xcolor}
\usepackage[colorlinks=true,urlcolor=blue,citecolor=blue,linkcolor=blue]{hyperref}
\hypersetup{breaklinks=true}
\usepackage{tcolorbox}
\usepackage{soul}

\newcommand{\be}{\begin{equation}}
\newcommand{\ee}{\end{equation}}

\newcommand{\bea}{\begin{eqnarray}}
\newcommand{\eea}{\end{eqnarray}}

\renewcommand{\Re}{{\rm \, Re\,}}

\renewcommand{\vec}[1]{{\bf #1}}
\renewcommand{\phi}{\varphi}
\renewcommand{\epsilon}{\varepsilon}

\newcommand{\df}[1]{\,\delta{\left(#1\right)}}
\newcommand{\tf}[1]{\,\theta{\left(#1\right)}}
\newcommand{\sign}[1]{\,\mbox{sgn}\left({#1}\right)}

\begin{document}

\title{Bound states, resonances, and their thermodynamic properties in pseudospin-1 systems with short-range impurities}
\date{August 31, 2026}

\author{E. V. Gorbar}
\affiliation{Faculty of Physics, Kyiv National Taras Shevchenko University, 64/13 Volodymyrska st., 01601 Kyiv, Ukraine}
\affiliation{Bogolyubov Institute for Theoretical Physics, 14-b Metrolohichna st., 03143 Kyiv, Ukraine}
\thanks{All authors contributed equally}

\author{Pavlo~Sukhachov}
\email{pavlo.sukhachov@missouri.edu}
\affiliation{Department of Physics and Astronomy, University of Missouri, Columbia, Missouri, 65211, USA}
\affiliation{MU Materials Science \& Engineering Institute, University of Missouri, Columbia, Missouri, 65211, USA}
\thanks{All authors contributed equally}

\begin{abstract}
Bound states and resonances induced by short-range impurities modeled by circular potential wells are analyzed in the vicinity of flat and dispersive bands in gapped and gapless pseudospin-1 systems. We find that the bound and resonant states derived from the flat band show unusual characteristics originating from the multicomponent structure of pseudospin-1 fermions, which are distinct from those for pseudospin-$\tfrac{1}{2}$ fermions. Contrary to gapped Dirac systems and unlike bound states in the vicinity of the upper dispersive band, the bound states derived from the flat band occur for any value of the total angular momentum. The energies of these bound states with higher angular momentum $j$ tend to decrease with $|j|$. In addition, it is found that their wave functions are localized at the potential well edge and the localization increases with $|j|$. The signatures of the impurity states in the local density of states are determined. Using the Anderson model for independent electrons in the disorder potential, the thermodynamic potential, entropy density, and heat capacity are obtained. In the regime dominated by bound states derived from the flat band, the entropy density monotonically increases with temperature and saturates, whereas the heat capacity exhibits a single maximum.
\end{abstract}

\maketitle

\section{Introduction}

Recent advances in topological and Dirac materials have stimulated considerable interest in quasiparticles with unconventional band structures. Among these systems, pseudospin-1 fermions constitute a particularly intriguing class of quasiparticles characterized by an energy spectrum consisting of two linearly dispersing bands intersected by a dispersionless flat band at the band-crossing point \cite{Sutherland1986, Dora2011}. Such a band structure emerges in a variety of platforms, including dice, $\alpha-T_3$, and Lieb lattices~\cite{Bercioux2009, Shen2010, Graf-Piechon-DesigningFlatbandTightbinding-2021}. These lattices can be realized in photonic crystals, artificial electronic lattices, and engineered optical lattices~\cite{Taie-Takahashi-CoherentDrivingFreezing-2015, Slot-Swart-ExperimentalRealizationCharacterization-2016, Leykam2018, Lebrat-Greiner-FerrimagnetismUltracoldFermions-2026}. Recently, pseudospin-1 fermions were observed in the van der Waals metal Pd$_5$AlI$_2$~\cite{Devarakonda-Roy-FrustratedElectronHopping-2025}, which realizes a decorated chequerboard model. 

The presence of a flat band plays a crucial role in determining the physical behavior of pseudospin-1 systems. Because the density of states associated with the flat band is strongly enhanced near the band-touching point, many-body correlations~\cite{Iurov-Huang:2020, Soni-Dagotto-FlatBandsFerrimagnetic-2020, Gorbar-Oriekhov:2021}, localization effects~\cite{Singh-Sharma-QuantumInterferencePseudospin1-2023}, and impurity scattering~\cite{Vigh2013, Xu-Lai-RevivalResonantScattering-2016} may become significantly altered~\cite{Green2010}. Consequently, thermodynamic quantities such as the specific heat, entropy, compressibility, and magnetic susceptibility can exhibit unconventional temperature and chemical-potential dependencies~\cite{Raoux2014, Horing-Mancini-GreensFunctionAnalysis-2021, Lebrat-Greiner-FerrimagnetismUltracoldFermions-2026}. Understanding these properties is essential for clarifying the interplay between band topology, disorder, and thermal fluctuations in pseudospin-1 materials.

In realistic systems, disorder and impurities are unavoidable and can strongly influence low-energy excitations~\cite{Balatsky-Zhu:rev-2006, DasSarma-Rossi:review-2011}. In particular, short-range impurities constitute an important source of elastic scattering in two-dimensional Dirac-like materials. Unlike long-range Coulomb disorder, short-range impurity potentials lead to momentum-independent scattering that substantially modifies the quasiparticle lifetime and density of states near the Dirac point. Furthermore, impurities allow for bound states, which exist in the energy gap, and resonances, which overlap with continuum states. Due to the low density of states near Dirac points, the latter are particularly relevant for Dirac systems~\cite{Balatsky-Zhu:rev-2006}. In pseudospin-1 systems, disorder strongly modifies the singular flat-band contribution to the low-energy density of states and produces distinctive localization behavior~\cite{Vigh2013, Singh-Sharma-QuantumInterferencePseudospin1-2023}. Consequently, thermodynamic observables derived from the density of states are expected to be particularly sensitive to impurity-induced broadening. Thus, investigating the role of short-range impurities is necessary for developing a comprehensive theoretical description of experimentally accessible pseudospin-1 materials.

Previous studies have extensively explored transport and optical responses of pseudospin-1 fermions~\cite{Illes2016, Kovacs-Cserti:2016, Carbotte-Nicol:2019, Iurov-Huang:2020, Han-Lai:2022, Oriekhov-Gusynin:2022, Iurov-Huang-OpticalConductivityGapped-2023,SOG:part2-2023}, revealing phenomena such as super-Klein tunneling, unconventional Landau levels~\cite{Balassis-Roslyak:2019}, and anomalous conductivity~\cite{Bercioux2009, Vigh2013, Illes2015, Biswas2016}. Thermodynamic properties of clean dice and Lieb lattices were studied in Refs.~\cite{Raoux2014, Oriekhov-Loktev:2020, Horing-Mancini-GreensFunctionAnalysis-2021, Kulynych-Oriekhov-DifferentialEntropyParticle-2022, Lara-DaCosta-InterconvertibilityThermodynamicProperties-2026}. The formation and properties of the impurity-induced bound states were investigated in Refs.~\cite{Gorbar-Oriekhov:2018, Han-Lai-AtomicCollapsePseudospin1-2019, Wang-Peeters:2021}. However, comparatively less attention has been devoted to the thermodynamic behavior of these systems in the presence of disorder. In particular, the combined influence of flat-band physics and impurity-induced bound states on equilibrium thermodynamic quantities remains insufficiently understood. Addressing this problem is important not only from a fundamental perspective but also for potential applications in tunable electronic and thermal devices based on engineered lattice systems.

In this work, we revisit the properties of the bound states and resonances in pseudospin-1 systems with short-range disorder. By considering a finite potential well model and combining analytical and numerical approaches, we show that the bound states for pseudospin-1 systems, which decouple from the flat band, exist for an arbitrary value of the angular momentum $j$ and, contrary to the case of gapped Dirac materials, their localization at the boundary of the potential well increases with $|j|$. The mathematical reason for this unique feature is the singular character of the pseudospin-related interaction term in an effective one-component equation for pseudospin-1 quasiparticles at the turning point. Assuming the Anderson model of disorder, we investigate the thermodynamic properties of the bound states as a function of temperature and chemical potential, showing that these properties resemble those of systems with a finite number of available energy states. In addition to the bound states, we also addressed the properties of resonances in gapless pseudospin-1 systems, demonstrating that, like the bound states, these resonances possess unusual characteristics.

The paper is organized as follows. In Sec.~\ref{sec:model}, we introduce the gapped model Hamiltonian for pseudospin-1 fermions and consider bound states in the gapped dice model for a circular potential well. The corresponding bound state energies are found in Sec.~\ref{sec:energy-levels}. Using the Anderson model of disorder, the thermodynamic potential, pressure, isothermal compressibility, and heat capacity are calculated in Sec.~\ref{sec:thermodynamics}. Sec.~\ref{sec:gapless} presents the calculation of the impurity resonances, scattering phases, and the local density of states (LDOS) in the gapless dice model. Finally, Sec.~\ref{sec:conclusion} summarizes our main results and outlines possible directions for future research. Technical details are summarized in Appendices~\ref{sec:App-Bessel}--\ref{sec:tmd}.

\section{Model and bound states}
\label{sec:model}

\begin{figure}[!ht]
\centering
\includegraphics[width=0.45\textwidth]{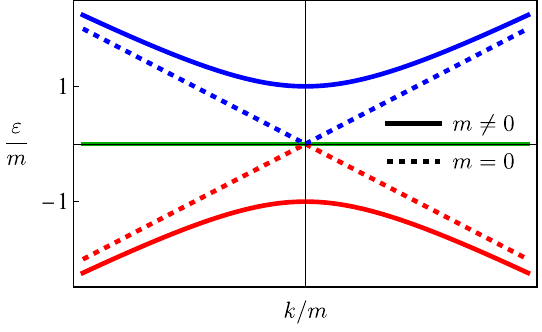}
\caption{
Energy bands defined by Hamiltonian~\eqref{Hamiltonian-free} for $m\neq0$ (solid lines) and $m=0$ (dashed lines).
}
\label{fig:energy}
\end{figure}

We focus on 2D pseudospin-1 fermions. The simplest isotropic Hamiltonian for such fermions in momentum space reads as
\begin{equation}
\label{model-H-def}
H_0(\mathbf{k}) = S_xk_x + S_yk_y + S_zm,
\end{equation}
where $\bm{S}$ are spin-1 matrices and we set $\hbar=v_F=1$. In our calculations, we fix the following explicit form of the Hamiltonian
\begin{eqnarray}
H_0(\vec{k}) &=& \frac{1}{\sqrt{2}} \left(\begin{array}{ccc}
0 & k_x-ik_y & 0\\
k_x+ik_y & 0 & k_x-ik_y\\
0 & k_x+ik_y & 0
\end{array}\right) \nonumber\\
&+& m \left(\begin{array}{ccc}
1 & 0 & 0\\
0 & 0 & 0\\
0 & 0 & -1
\end{array}\right).
\label{Hamiltonian-free}
\end{eqnarray}
Note that this form of Hamiltonian corresponds to a low-energy Hamiltonian of the K valley in a dice model~\cite{Bercioux2009}. The opposite valley is related by time-reversal symmetry. This Hamiltonian describes pseudospin-1 fermions whose energy bands consist of two dispersive bands $\epsilon=\pm\sqrt{\mathbf{k}^2+m^2}$ and one completely flat band $\epsilon=0$. The dispersive bands are separated from the flat band by the gap $m$. In a gapless dice model with $m=0$, all three bands intersect at $\mathbf{k}=0$. We plot the energy spectrum of Hamiltonian~\eqref{Hamiltonian-free} in Fig.~\ref{fig:energy}.

In the presence of an impurity, translation invariance is lost and one should replace $\mathbf{k}\to -i\bm{\nabla}$. For pseudospin-1 fermions, impurities could be of a scalar type perturbing the Hamiltonian as $H=H_0+V(\mathbf{r})I_3$ with $I_3$ being the unit matrix, or of a magnetic type, which couples impurity spin to pseudospin. 

In our study, we consider scalar (electrostatic) impurities, which could correspond to charged defects, adsorbed atoms, and gate- or tip-induced potentials. We model screened charged defects with a short-range potential as a circular potential well of depth $V_0$ and radius $r_0$. The corresponding potential is $V(\mathbf{r}) = -V_0 \theta{(r_0-r)}$. This model provides a regularization for a commonly used $\delta$-function potential, which is known to lead to ambiguities for relativistic (more generally, systems of coupled) equations~\cite{Sutherland-Mattis-AmbiguitiesRelativisticFunction-1981}. Yet, as we will demonstrate, the sharp boundary of $V(\mathbf{r}) = -V_0 \theta{(r_0-r)}$ may also cause difficulties, see Sec.~\ref{sec:regularized}.

The bound states for impurities of the scalar type were studied in Ref.~\cite{Gorbar-Oriekhov:2018}. In what follows, we reproduce the key expressions and provide a more detailed analytical analysis of bound state energies, paying special attention to their dependence on the angular momentum as well as to the localization properties of bound-state wave functions.

The quasiparticle wave function of a state with total angular momentum $j$ in the gapped pseudospin-1 model with a circular potential well defined by $V(r)= -V_0 \theta{(r_0-r)}$ is given by~\cite{Gorbar-Oriekhov:2018}
\begin{eqnarray}
\Psi = \frac{1}{r}\left(\begin{array}{c}
a(r)e^{i(j-1)\phi}\\
ic(r)e^{ij\phi}\\
b(r)e^{i(j+1)\phi}
\end{array}
\right).
\end{eqnarray}
Here,
\begin{eqnarray}
\frac{c(r)}{r} &=& A\theta(r_0-r)\,J_{|j|}(v_1(\epsilon+V_0)r) \nonumber\\
&+& \frac{AJ_{|j|}(v_1(\epsilon+V_0)r_0)}{K_{|j|}(v_2(\epsilon)r_0)}\theta(r-r_0)\,K_{|j|}(v_2(\epsilon)r),
\label{wave-function}
\end{eqnarray}
where $v_1(\varepsilon)=\sqrt{\varepsilon^2-m^2}$,  $v_2(\varepsilon)=\sqrt{m^2-\varepsilon^2}$, and the wave function components $a(r)$ and $b(r)$ are expressed through its component $c(r)$
\begin{equation}
\label{components}
\begin{aligned}
a(r) &= \frac{1}{\sqrt{2}} \frac{c^{\prime}(r) +\frac{j-1}{r}c(r)}{\epsilon-V(r)-m}, \\
b(r) &= \frac{1}{\sqrt{2}} \frac{c^{\prime}(r) -\frac{j+1}{r}c(r)}{\epsilon-V(r)+m}.
\end{aligned}
\end{equation}

The bound state energies are defined by the following equation:
\begin{widetext}
\begin{eqnarray}
&&\frac{v_1(\epsilon+V_0)}{m-\epsilon-V_0}\frac{{\rm sgn}(j)J_{|j|-\rm{sgn}(j)}(v_1(\epsilon+V_0)r_0)}{J_{|j|}(v_1 (\epsilon+V_0)r_0)}
+\frac{v_1(\epsilon+V_0)}{m+\epsilon+V_0}\frac{{\rm sgn}(j) J_{|j|+\rm{sgn}(j)}(v_1(\epsilon+V_0)r_0)}{J_{|j|}(v_1 (\epsilon+V_0)r_0)}=\nonumber\\
&&-\frac{v_2(\epsilon)}{m-\epsilon}\frac{K_{|j|-\rm{sgn}(j)}(v_2(\epsilon)r_0)}{K_{|j|}(v_2(\epsilon)r_0)}
+\frac{v_2(\epsilon)}{m+\epsilon}\frac{K_{|j|+\rm{sgn}(j)}(v_2(\epsilon)r_0)}{K_{|j|}(v_2(\epsilon)r_0)},
\label{spectrum-1}
\end{eqnarray}
\end{widetext}
where we define $\mbox{sign}(0)=1$ for $j=0$. 

Bound states are well defined only in the energy gap, hence, we assume $|\epsilon| < m$. In addition, since the presence of the flat band is a hallmark feature of pseudospin-1 fermions, we pay special attention to the bound states in the vicinity of the flat band $V_0 \ll m$. Therefore, we focus on $|\epsilon +V_0| < m$.

\section{Analysis of bound states in gapped pseudospin-1 model}
\label{sec:energy-levels}

In this section, we first analyze the characteristic equation \eqref{spectrum-1} in the vicinity of the flat and upper dispersive bands. Then, we support our analytical results with numerical calculations. Finally, we discuss the characteristics of wave functions of bound states and their localization properties.

\subsection{Bound states in vicinity of the flat band}
\label{sec:vicinity-flat-band}

Assuming $|\epsilon| < m$ and $|\epsilon +V_0| < m$ as well as using $J_{\alpha}(ix)=i^{\alpha}I_{\alpha}(x)$, we find that Eq.~(\ref{spectrum-1}) takes the form
\begin{widetext}
\begin{eqnarray}
&&\frac{\tilde{v}_1(\epsilon+V_0)}{m-\epsilon-V_0}\frac{I_{|j|-\rm{sgn}(j)}(\tilde{v}_1(\epsilon+V_0)r_0)}{I_{|j|}(\tilde{v}_1 (\epsilon+V_0)r_0)}
-\frac{\tilde{v}_1(\epsilon+V_0)}{m+\epsilon+V_0}\frac{ I_{|j|+\rm{sgn}(j)}(\tilde{v}_1(\epsilon+V_0)r_0)}{I_{|j|}(\tilde{v}_1 (\epsilon+V_0)r_0)}=\nonumber\\
&&-\frac{v_2(\epsilon)}{m-\epsilon}\frac{K_{|j|-\rm{sgn}(j)}(v_2(\epsilon)r_0)}{K_{|j|}(v_2(\epsilon)r_0)}
+\frac{v_2(\epsilon)}{m+\epsilon}\frac{K_{|j|+\rm{sgn}(j)}(v_2(\epsilon)r_0)}{K_{|j|}(v_2(\epsilon)r_0)},
\label{spectrum-3}
\end{eqnarray}
\end{widetext}
where $\tilde{v}_1(\varepsilon+V_0)=\sqrt{m^2-(\varepsilon+V_0)^2}$. This equation possesses the following symmetry:
\begin{equation}
j \to -j,\quad \epsilon \to -\epsilon, \quad
V_0 \to -V_0.
\label{symmetry}
\end{equation}
Therefore, we fix $j \geq 0$ and derive the solutions with negative $j$ by using Eq.~\eqref{symmetry}, i.e., replacing the signs of $\epsilon$ and $V_0$ in the corresponding analytical expressions.

Using the standard relations for the Bessel functions, see Appendix~\ref{sec:App-Bessel}, Eq.~(\ref{spectrum-3}) can be equivalently rewritten as follows:
\begin{widetext}
\begin{eqnarray}
&&\frac{\tilde{v}_1(\epsilon+V_0)}{m^2-(\epsilon+V_0)^2}\left(\frac{mj}{r_0\tilde{v}_1(\epsilon+V_0)}+(\epsilon+V_0)(\ln I_j(x))^{\prime}_{|_{x=r_0\tilde{v}_1(\epsilon+V_0)}} \right)\nonumber\\
&&=\frac{v_2(\epsilon)}{m^2-\epsilon^2}\left(\frac{mj}{r_0v_2(\epsilon)}+\epsilon(\ln K_j(x))^{\prime}_{|_{x=r_0v_2(\epsilon)}} \right).
\label{spectrum-4}
\end{eqnarray}
\end{widetext}

As was numerically shown in Ref.~\cite{Gorbar-Oriekhov:2018}, there are two types of bound states in the gapped pseudospin-1 system: states that decouple from the upper ($\varepsilon=m$) continuum and the states that decouple from the flat band. Let us start with the latter. Assuming $|\epsilon| \ll m$ and $|V_0| \ll m$ as well as approximating $\tilde{v}_1(\epsilon+V_0)\approx m$ and $v_2(\epsilon)\approx m$, Eq.~(\ref{spectrum-4}) takes the form
\begin{equation}
\epsilon=-V_0\frac{\left(\ln I_j(x_0)\right)^{\prime}}{(\ln I_j(x_0))^{\prime}-(\ln K_j(x_0))^{\prime}},\quad x_0=r_0m.
\label{energy-levels}
\end{equation}
To make an analytical advance, we consider a few simplifying cases, such as (i) narrow potential $mr_0 \ll1$, (ii) wide potential $mr_0\gg1$, and (iii) large $j$ limit.

\subsubsection{Narrow potential well and large $j$ limit}
\label{sec:energy-levels-1}

In the case of a narrow potential well, i.e., $mr_0\ll 1$, using the asymptotic behavior at small $x$ and $j>1$, see Eq.~\eqref{asymptotics-small}, and keeping only leading $(mr_0)^2$ terms in the Bessel functions, Eq.~(\ref{energy-levels}) acquires the form
\begin{eqnarray}
\epsilon=-V_0\frac{1+\frac{(mr_0)^2}{2j(j+1)}}{1+\frac{(mr_0)^2}{2j(j+1)}+1+\frac{(mr_0)^2}{2j(j-1)}}.
\label{spectrum-6}
\end{eqnarray}
For $mr_0 \to 0$, it is easy to find the corresponding solution $\epsilon=-V_0/2$. Representing $\epsilon=-V_0/2+\delta\epsilon$ and determining $\delta\epsilon$ to leading order in $(mr_0)^2$, we obtain the energy of a state with angular momentum $j >1$
\begin{equation}
\epsilon_{j>1}(V_0) =-\frac{V_0}{2}\left[1-\frac{(mr_0)^2}{2j(j^2-1)}\right].
\label{solution-narrow}
\end{equation}
The cases $j=0$ and $j=1$ require separate consideration.

In the case $j=0$, using Eq.~\eqref{energy-asym} in Eq.~\eqref{energy-levels}, we obtain the approximate solution
\begin{equation}
\epsilon_0(V_0) =\frac{V_0}{2} (mr_0)^2 \ln{\left(\frac{mr_0}{2}\right)}.
\label{solution-special}
\end{equation}
For the angular momentum $j=1$, using the asymptotics in Eq.~\eqref{energy-asym}, we find the following solution to Eq.~(\ref{energy-levels}):
\begin{equation}
\epsilon_1(V_0) = -\frac{V_0}{2} \left[1 +\frac{(mr_0)^2}{2} \ln{\left(\frac{mr_0}{2}\right)} \right].
\label{solution-special-j=1}
\end{equation}

In the limit of large $j$, the asymptotics of modified Bessel functions $K_j(x)$ and $I_j(x)$ are given in Eq.~\eqref{app-asymptotics-large-j}. The first derivatives of the logarithm of these asymptotics for $j \gg 1$ have the same form as the first derivatives of the logarithm of the asymptotics in Eq.~\eqref{asymptotics-small}, therefore, we obtain the same energy levels defined in Eq.~\eqref{solution-narrow}.

It is also worth noting that, according to Eq.~\eqref{solution-narrow}, the energy of states with higher $j$ is {\it lower} and approaches $-V_0/2$. This is a rather counterintuitive result. Indeed, intuition developed in standard quantum mechanical problems suggests that a higher angular momentum $|j|$ should correspond to a higher potential barrier and, as a result, bound state energy should increase with $|j|$. Another puzzling observation is the tendency of energy levels with higher $j$ to approach $-V_0/2$.

\subsubsection{Regularized potential well}
\label{sec:regularized}

To gain an insight into these peculiar behaviors of the bound states, we consider a regularized potential with continuous derivatives,
\begin{equation}
V(x)=\frac{V_0}{2}\left\{\tanh{\left[\lambda\left(\frac{x}{x_0}-1\right)\right]} -1 \right\},
\label{regularized}
\end{equation}
which reproduces the potential well $V(x)= -V_0 \tf{x_0-x}$ at $\lambda\to +\infty$. According to Eq.~(\ref{components}), the components $a(r)$ and $b(r)$ of the wave function are expressed through the component $c(r)$. Therefore, the system of differential equations reduces to a \emph{second order} differential equation for $f(r)=c(r)/r$
\begin{eqnarray}
&&f^{\prime\prime}+f^{\prime}\left[\frac{1}{r}+\frac{V^{\prime}}{\epsilon-V}\frac{(\epsilon-V)^2+m^2}{(\epsilon-V)^2-m^2}\right] \nonumber\\
&&+f\left[(\epsilon-V)^2-m^2-\frac{j^2}{r^2}+\frac{2j}{r}\frac{mV^{\prime}}{(\epsilon-V)^2-m^2}\right]=0,\nonumber\\
\label{equation-component-f-1}
\end{eqnarray}
which is convenient for the analysis of the regularized continuous potential given in Eq.~(\ref{regularized}). The reason why a system of three 1st-order differential equations for pseudospin-1 fermions reduces to a second (rather than third) order differential equation is the linear relation among its three components~\cite{Gorbar-Oriekhov:2018}
\begin{equation}
\frac{\sqrt{2}j}{r}c(r) +[m-\epsilon+V(r)]a(r)+[m+\epsilon-V(r)]b(r)=0.
\end{equation}

The key difference of Eq.~(\ref{equation-component-f-1}) for gapped pseudospin-1 fermions from a similar effective one-component equation in gapped graphene is the singularity of the interaction term with $f^{\prime}V^{\prime}/(\epsilon-V)$ at the turning point $x_t$ where $\epsilon=V(x_t)$. This singularity arises due to the multicomponent nature of pseudospin-1 fermions. Indeed, according to the analysis in Appendix~\ref{sec:tmd}, the analog of Eq.~\eqref{equation-component-f-1} in gapped graphene at zero magnetic field with gap $\Delta$ reads
\begin{eqnarray}
&&f^{\prime\prime}+\frac{f^{\prime}V^{\prime}}{\epsilon-V-\Delta} \nonumber\\
&&-\left[\frac{(j-1)^2-1/4}{r^2}-\frac{(j-\frac{1}{2})V^{\prime}}{r(\epsilon-V-\Delta)}\right]f \nonumber\\
&&= -\frac{(\epsilon-V)^2-\Delta^2}{v^2_F}f,
\label{equation-generic-1-main}
\end{eqnarray}
where $f(r)$ determines the wavefunction amplitude, see Eq.~\eqref{tmd-psi} and the text below it. Although a similar term is present in Eq.~(\ref{equation-generic-1-main}), the interaction term $f^{\prime}V^{\prime}/(\epsilon -V -\Delta)$ connected with the pseudospin-$\tfrac{1}{2}$ character of electron quasiparticles in graphene is suppressed by gap $\Delta$ and, hence, the turning point is not a singular point of Eq.~(\ref{equation-generic-1-main}). It is worth recalling that the turning point is a regular point for the Schr\"{o}dinger equation too and its spin-orbit interaction term, proportional to the product of the first derivatives of potential and wave function, is not singular. As we show in Sec.~\ref{sec:gapped-wf}, the singularity of the $f^{\prime}V^{\prime}/(\epsilon-V)$ interaction term of Eq.~(\ref{equation-component-f-1}) is responsible for a strong localization of bound-state wave functions at the turning point.

The analysis of Eq.~(\ref{equation-component-f-1}) in Appendix \ref{sec:App-1} shows that, in the limit $\lambda\to\infty$ and the leading nontrivial order in $x_0$ and $1/\lambda$, bound-state energies are given by
\begin{equation}
\label{energy-sol-a-x0}
\epsilon_{j>1}(V_0) \approx -\frac{V_0}{2} \left[1 - \frac{V_0}{2m\lambda} -\frac{x_0^2}{2j(j^2-1)} 
\right].
\end{equation}
In the limit $\lambda\to \infty$, these energies tend to bound-state energies given in Eq.~(\ref{solution-narrow}) for the unregularized potential. In addition, bound-state energies decrease with $j$ at large $j$, and the presence of the term $1/\lambda$ shows that the energies do not tend to $-V_0/2$ at $j\to \infty$ as this value is affected by $1/\lambda$, i.e., by the form of the potential~\footnote{The correction to $-V_0/2$ in Eq.~\eqref{energy-sol-a-x0} is, however, vanishingly small at $\lambda\to\infty$ and $V_0\ll m$.}.

\subsubsection{Wide potential well}
\label{sec:energy-levels-2}

For a wide potential well, i.e., $mr_0\gg 1$, using the asymptotics in Eq.~\eqref{function-asymptotics}, we find that Eq.~\eqref{energy-levels} gives
\begin{equation}
 \epsilon_j(V_0) =-\frac{V_0}{2}\left\{1-\frac{1}{2mr_0}+\frac{3(4j^2-1)}{16(mr_0)^3}\right\}.
 \label{spectrum-wide-4}
\end{equation}
As in the case of a narrow potential well considered in Sec.~\ref{sec:energy-levels-1}, the energy of bound states decreases with $j$. Therefore, this unusual behaviour is not an artifact of the narrow potential well approximation; later, we confirm this observation via numerical calculations.

\subsection{Bound states in vicinity of upper band}
\label{sec:energy-levels-upper}

For completeness and to demonstrate the realization of different types of impurity-induced states for pseudospin-1 fermions, we also address bound states in the vicinity of the dispersive bands.

In the vicinity of the upper band, assuming $|V(r)| \ll m$ and $\epsilon=m+E$ with $E<0$ and $|E|\ll m$, Eq.~(\ref{equation-component-f-1}) acquires the form
\begin{equation}
f^{\prime\prime}+\frac{f^{\prime}}{r}-\frac{j^2f}{r^2}+\left(f^{\prime} +\frac{jf}{r} \right)\frac{V^{\prime}}{E-V} +2m(E-V)f=0.
\label{equation-component-c-upper-1}
\end{equation}
Note that, similar to Eq.~(\ref{equation-component-f-1}), which describes bound states in the vicinity of the flat band, Eq.~(\ref{equation-component-c-upper-1}) is singular at the turning point $x_t$ where $V(x_t)=E$.

For the considered potential well and $r<r_0$, the above equation reduces to the canonical 2D Schr\"{o}dinger equation in a central potential
\begin{equation}
f^{\prime\prime}+\frac{f^{\prime}}{r}-\frac{j^2f}{r^2}+2m(E+V_0)f=0.
\label{equation-component-c-upper-2}
\end{equation}
For $r>r_0$, we have the same equation but with $V_0=0$. Regular at the origin solution to Eq.~(\ref{equation-component-c-upper-2}) is given by $f_I(r)=C_1J_j(\sqrt{2m(E+V_0)}\,r)$. Further, $f_{II}(r)=C_2K_j(\sqrt{-2mE}\,r)$ is a normalizable solution for $r>r_0$, where we assume $|E|<V_0$.

Matching these solutions at $r_0$, we find that $C_2=C_1J_j(\sqrt{2m(E+V_0)}\,r_0)/K_j(\sqrt{-2mE}\,r_0)$. Since $V^{\prime}(r)=V_0\delta(r-r_0)$, we determine the second matching condition by integrating Eq.~(\ref{equation-component-c-upper-1}) across $r_0$,
\begin{eqnarray}
&&\int \left[f^{\prime\prime}\frac{1}{E-V}+\frac{f^{\prime}V^{\prime}}{(E-V)^2}+\frac{jf}{r}\frac{V^{\prime}}{(E-V)^2}\right]dr \nonumber\\
&&=\int \left[\left(f^{\prime}\frac{1}{E-V}\right)^{\prime}+\frac{jf}{r}\frac{V^{\prime}}{(E-V)^2}\right]dr.
\end{eqnarray}
We find the second matching condition
\begin{equation}
f^{\prime}_{II}(r_0)(E+V_0)-f^{\prime}_I(r_0)E+\frac{jf(r_0)}{r_0}V_0=0.
\end{equation}

Using the matching conditions, we obtain the following characteristic equation:
\begin{eqnarray}
&&(E+V_0)\left(\ln K_j(\sqrt{-2mE}\,r)\right)^{\prime}_{|_{r=r_0}} \nonumber\\
&&-E\left(\ln J_j(\sqrt{2m(E+V_0)}\,r)\right)^{\prime}_{|_{r=r_0}}+\frac{jV_0}{r_0}=0.
\label{upper-band}
\end{eqnarray}
Note that this characteristic equation follows also from Eq.~(\ref{spectrum-4}) using $\tilde{v}_1(\epsilon+V_0)\approx i\sqrt{2m(E+V_0)}$, $v_2(\epsilon)\approx \sqrt{-2mE}$, $m^2-(\epsilon+V_0)^2\approx -2m(E+V_0)$, and $m^2-\epsilon^2\approx -2mE$.

To contrast the bound states that decouple from the upper band and the flat band, let us determine energies of bound states in the vicinity of the upper band in a narrow potential well when $mr_0 \to 0$. Using the asymptotics of $K_j(x)$ and $J_j(x)$ at small $x$ and $j>1$ given in Eq.~\eqref{asymptotics-small}, Eq.~(\ref{upper-band}) yields
\begin{eqnarray}
&&(E+V_0)\left(\frac{j}{r_0}-\frac{Emr_0}{j-1}\right)+E\left(\frac{j}{r_0}-\frac{(E+V_0)mr_0}{j+1}\right) \nonumber\\
&&-\frac{jV_0}{r_0}=0,
\label{upper-band-large}
\end{eqnarray}
which gives
\begin{equation}
E= -V_0 +\frac{j^2-1}{mr^2_0}.
\label{upper-band-solution}
\end{equation}
Note that bound-state energies $E$ increase with $|j|$. In addition, for a given $V_0$, there is no solution if $|j|>\sqrt{1+mr^2_0V_0}$ because $E$ should be negative for a bound state.

\subsection{Numerical analysis}
\label{sec:numerical}

\begin{figure*}[!ht]
\centering
\subfigure[]{\includegraphics[width=0.45\textwidth]{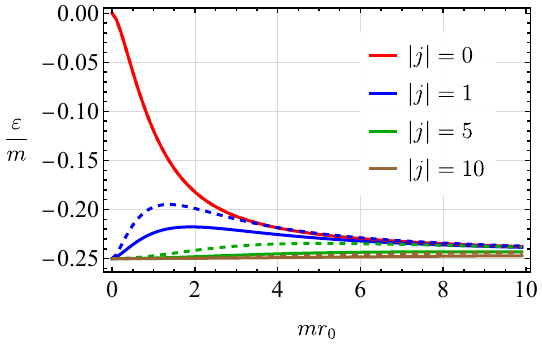}}
\subfigure[]{\includegraphics[width=0.435\textwidth]{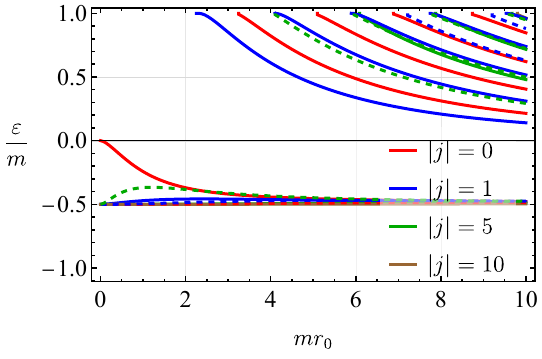}}
\caption{
Energies of bound states that split off from the flat band at $V_0/m=0.5$ (panel (a)) and energies of all bound states at $V_0/m=1$ (panel (b)) as a function of $mr_0$ for a few values of $j$. Solid and dashed lines correspond to $j \geq0$ and $j<0$, respectively.
}
\label{fig:levels}
\end{figure*}

We show numerical results for the dependence of $\epsilon/m$ on $mr_0$ and several values of the total angular momentum $j$ in Fig.~\ref{fig:levels}. These results are obtained by numerically solving Eq.~\eqref{spectrum-1}.

As one can see from Fig.~\ref{fig:levels}(a), the energy of the bound state with $j=0$, which decouples from the flat band, tends to zero for $mr_0 \to 0$, which is in agreement with Eq.~(\ref{solution-special}). The energies of other states with $|j|>0$ tend to $-V_0/2$ according to Eqs.~(\ref{solution-special-j=1}) and (\ref{solution-narrow}). Note that energies of bound states with $j$ and $-j$ are quite similar and differ notably only for $mr_0 \sim 1$. While the energy of the state with $j=0$ decreases monotonically with $mr_0$, the energies of the states with $|j|>0$ depend on $mr_0$ nonmonotonically. For $mr_0 \gg 1$, the energies of all states that decouple from the flat band tend to $-V_0/2$ in agreement with Eq.~(\ref{spectrum-wide-4}). The energies of the bound states that originate from the flat band decrease as $|j|$ grows, confirming our analytic results obtained in Sec.~\ref{sec:vicinity-flat-band} in the limit of large $j$. In addition and rather counterintuitively, energies of these states \emph{decrease} for larger $mr_0$ in agreement with the analytic results in Eq.~(\ref{spectrum-wide-4}).

Energies of bound states which decouple from the flat and upper dispersive band are plotted in Fig.~\ref{fig:levels}(b) as a function of $mr_0$ at $V_0/m = 1$. One can see that energies of bound states which decouple from the upper band monotonically decrease with $mr_0$ in contrast to energies of bound states which decouple from the flat band discussed above.

\subsection{Wave function localization}
\label{sec:gapped-wf}

To finalize this section, let us address the spatial localization of the bound states in the impurity potential. As we will show below, despite a different dependence of energies on the angular quantum number $j$, the bound states that decouple from the flat and dispersive bands show rather similar spatial localization properties.

\begin{figure*}[!ht]
\centering
\subfigure[]{\includegraphics[width=0.42\textwidth]{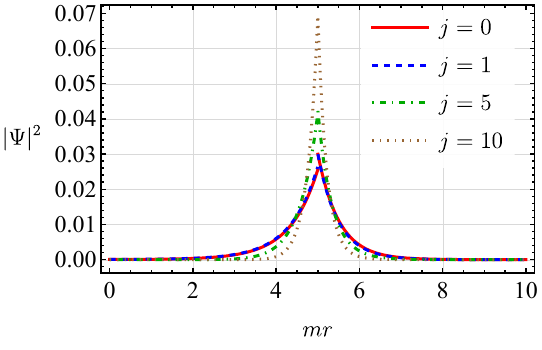}}
\subfigure[]{\includegraphics[width=0.42\textwidth]{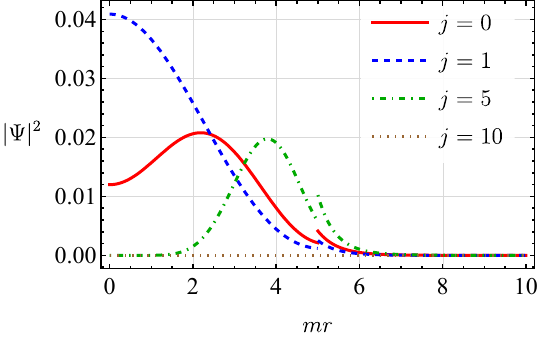}}
\caption{The probability density of a bound state wave function $|\Psi|^2$ as a function of $mr$ for fixed $mr_0=5$, where $|\Psi|^2$ is normalized to unity by integrating it over the whole space. Panels (a) and (b) correspond to the bound states that decouple from the flat band (panel (a)) and the lowest-energy bound states that decouple from the upper band (panel (b)).
}
\label{fig:psi2}
\end{figure*}

The large-$j$ asymptotics of functions $I_j(x)$ and $K_j(x)$ given in Eq.~(\ref{app-asymptotics-large-j}) suggest that the wave functions of bound states in the vicinity of the flat band are concentrated at the potential well edge, $r=r_0$. To confirm and illustrate this, we plot the normalized probability density $|\Psi|^2$ in Fig.~\ref{fig:psi2}(a). Here, $|\Psi|^2$ is plotted as a function of $mr$ and is normalized according to $\int d^2r\,|\Psi(\mathbf{r})|^2=1$. The components of the wave function are given in Eqs.~(\ref{wave-function}) and (\ref{components}).

The results in Fig.~\ref{fig:psi2}(a) clearly show that the localization of bound state wave functions at $r_0$ essentially {\it increases} with $j$, in agreement with the large-$j$ asymptotics given in Eq.~(\ref{app-asymptotics-large-j}). It is instructive to discuss our results in the context of Ref.~\cite{Pottelberge}, which argued that, since the flat band is dispersionless, bound state wave functions are localized at the turning point $x_t$ defined by $\epsilon=V(x_t)$. The mathematical reason is connected with the fact that the turning point is a singular point of the corresponding effective one-component differential equation. For a sharp potential well, the turning point is $r_0$ for all bound states with $-V_0<\epsilon<0$, therefore, our results agree with the arguments given in Ref.~\cite{Pottelberge}. For a smooth regularization of the square well, the turning point lies within the transition region and approaches $r_0$ in the sharp-boundary limit.

In view of such a strong localization of the wave function of bound states at the turning point and the existence of bound states for any large $|j|$, it is worth recalling the localization properties of bound states in gapped graphene. For massive Dirac quasiparticles in a circular potential well, there is a critical value of the total angular momentum above which bound states are absent. Indeed, for bound states near the upper band, one can use the 2D Schr\"{o}dinger equation, for which the mathematical bounds for the number of bound states and the critical values of the total angular momentum are well established~\cite{Chadan}.

Another feature of the probability density $|\Psi|^2$ in Fig.~\ref{fig:psi2}(a) is the presence of jumps of the probability density at $r=r_0$. This jump originates from components $a(r)$ and/or $b(r)$ of the wave function \eqref{wave-function}, which are determined by discontinuous $c^{\prime}(r)$. Only their sum, $a(r)+b(r)$, is continuous at $r_0$ as required by the matching conditions derived in Ref.~\cite{Gorbar-Oriekhov:2018}. A mathematical reason for the presence of these jumps is connected with the fact that although the wave function for pseudospin-1 fermions has three components, only two of them are independent. Therefore, the corresponding system of three differential equations can be reduced to a single differential equation of the second (not third!) order~\cite{Gorbar-Oriekhov:2018} given by Eq.~(\ref{equation-component-f-1}). This fact, combined with the step-like potential, results in a jump of $|\Psi|^2$ at $r_0$.

To contrast the behavior of the bound states that decouple from the flat band and the upper band, we plot the probability density $|\Psi|^2$ for the latter in  Fig.~\ref{fig:psi2}(b). The wave functions are normalized to unity by integrating over the whole space. Comparing Figs.~\ref{fig:psi2}(a) and \ref{fig:psi2}(b), one can see that the wave functions of bound states in the vicinity of the upper band are not strongly localized at the potential well edge, although as $j$ increases, the wave functions are squeezed to the edge. Note that a similar behavior is realized for the wave functions of bound states in gapped graphene in a circular potential well~\cite{Giavaras-Nori-DiracGapinducedGraphene-2011}.

\section{Thermodynamic properties in gapped dice model}
\label{sec:thermodynamics}

In this section, we consider the thermodynamic properties of the bound states that decouple from the flat band. Our choice is motivated by their unusual properties and lower energies compared to their counterpart originating from the upper dispersive band. We use the Anderson model of disorder~\cite{Anderson} and assume that the impurity density is small, hence, bound states of different impurities can be treated as independent. In the Anderson model of disorder, depths $V_0$ of potential wells are distributed in the interval $[0,W]$.

The thermodynamic potential is defined as
\begin{equation}
\Omega=-\frac{N_iT}{W}\int^{W}_0dV_0\,\sum_j \ln\left[1+e^{\frac{\mu-\epsilon_j(V_0)}{T}} \right],
\label{thermodynamic-potential}
\end{equation}
where $\mu$ is the chemical potential, $T$ is temperature measured in energy units, and $N_i$ is the number of impurities. We assume that $W<m$ in order to be consistent with the analysis in the previous section where bound states were studied for $V_0< m$.

According to our analysis in Sec.~\ref{sec:vicinity-flat-band}, energies of impurity-induced bound states at large $|j|$ decrease with $|j|$ and tend to $-V_0/2$. This means that the sum over $j$ in Eq.~(\ref{thermodynamic-potential}) is, in fact, divergent and should be regularized. Since we work in a low-energy continuum model, a natural cut-off for $|j|$ is given by the condition that $|j|/r_0$ does not exceed the Brillouin zone size defined by $\pi/a_l$, where $a_l$ is the lattice constant. Applying this regularization, which, for consistency, implies the inequality $r_0>a_l$, the density of the thermodynamic potential (\ref{thermodynamic-potential}) acquires the form
\begin{equation}
\Omega_{V} =\frac{\Omega}{V} =
- \frac{n_iT}{W}\int^{W}_0dV_0\, \sum^{j_{\rm max}}_{j=-j_{\rm max}} \ln\left(1+e^{\frac{\mu-\epsilon_j(V_0)}{T}} \right),
\label{thermodynamic-potential-1}
\end{equation}
where $V$ is the volume (area) of the system and $j_{\rm max}$ is the cutoff. The latter can be provided by a lattice spacing $a_l$ as $j_{\rm max} = [\pi r_0/a_l]$.

Having determined the thermodynamic potential, we proceed to finding other thermodynamic quantities for impurity-induced bound states. 
Using the standard relation between the thermodynamic potential and the pressure, $\Omega=-pV$, rewriting Eq.~(\ref{thermodynamic-potential-1}) as
\begin{equation}
p = \frac{n_iT}{W}\int^{W}_0dV_0\,\sum^{j_{\rm max}}_{j=-j_{\rm max}} \ln\left(1+e^{\frac{\mu-\epsilon_j(V_0)}{T}} \right),
\label{equation-state}
\end{equation}
and using the expressions from Ref.~\cite{Huang}, we find the following expression for the isothermal compressibility:
\begin{equation}
\kappa_T=\frac{V}{N\left(\frac{\partial p}{\partial\mu}\right)_{T,V}}\left(\frac{\partial^2p}{\partial\mu^2}\right)_{T,V}=\frac{1}{\left(\frac{\partial p}{\partial\mu}\right)^2_{T,V}}\left(\frac{\partial^2p}{\partial\mu^2}\right)_{T,V},
\label{compressibility-1}
\end{equation}
where we used $N=V(\partial p/\partial\mu)_{T,V}$. Then, using Eq.~(\ref{equation-state}), we obtain
\begin{equation}
\kappa_T=\frac{W}{n_iT}\frac{
\int^{W}_0dV_0\, \sum^{j_{\rm max}}_{j=-j_{\rm max}} f_j(V_0) \left[1-f_j(V_0)\right]}{\left[\int^{W}_0dV_0\,\sum^{j_{\rm max}}_{j=-j_{\rm max}} f_j(V_0)\right]^2},
\label{compressibility-2}
\end{equation}
where $f_j(V_0)=1/\left[1+ e^{(\epsilon_j(V_0) -\mu)/T}\right]$ is the Fermi-Dirac distribution.

The entropy density is
\begin{eqnarray}
&&s = -\frac{\partial \Omega_{V}}{\partial T}
=-\frac{n_i}{W}\int^{W}_0dV_0\, \sum^{j_{\rm max}}_{j=-j_{\rm max}} \nonumber\\
&&\times \left\{f_j(V_0) \ln{f_j(V_0)} + \left[1 -f_j(V_0)\right] \ln{\left[1 -f_j(V_0)\right]}\right\}.\nonumber\\
\label{entropy-density}
\end{eqnarray}

The heat capacity (per unit volume at constant volume and chemical potential) is given by~\cite{Landau-t5}
\begin{eqnarray}
c_{V} &=& T\frac{\partial s}{\partial T} 
=\frac{n_i}{W}\int^{W}_0dV_0\, \sum^{j_{\rm max}}_{j=-j_{\rm max}}  \left[\frac{\epsilon_j(V_0) -\mu}{T}\right]^2 \nonumber\\ 
&\times& f_j(V_0)\left[1 -f_j(V_0)\right].
\label{td-heat}
\end{eqnarray}

\begin{figure*}[!ht]
\centering
\subfigure[]{\includegraphics[width=0.24\textwidth]{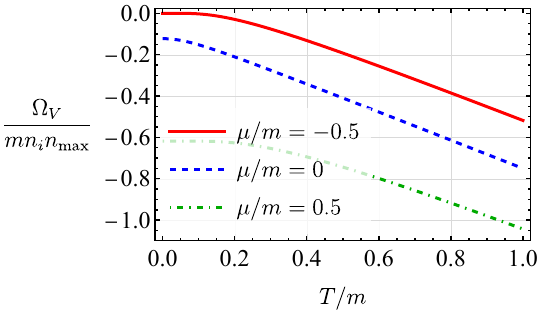}}
\subfigure[]{\includegraphics[width=0.24\textwidth]{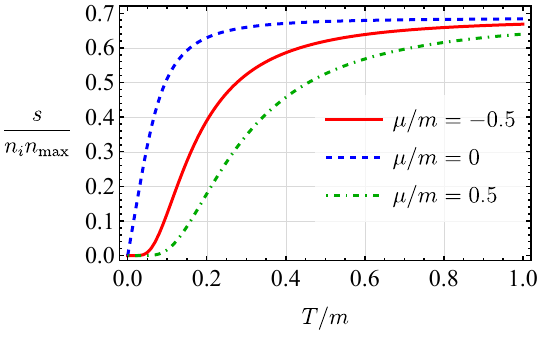}}
\subfigure[]{\includegraphics[width=0.24\textwidth]{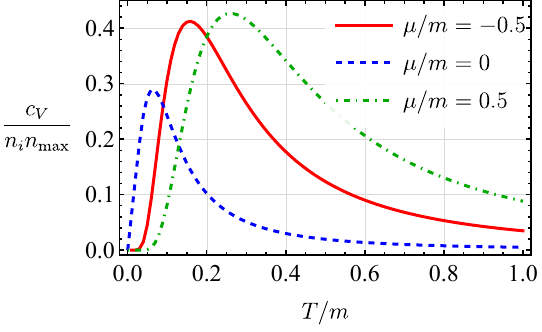}}
\subfigure[]{\includegraphics[width=0.24\textwidth]{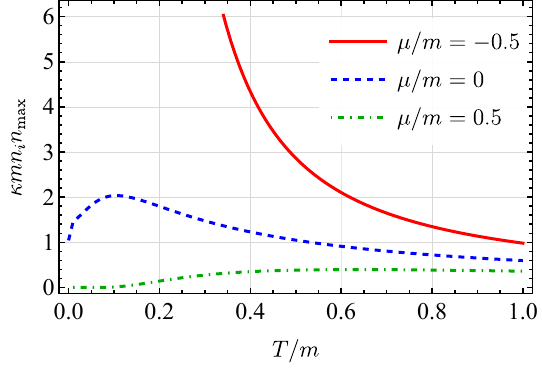}}
\caption{
The thermodynamic potential (panel (a)), entropy density (panel (b)), heat capacity (panel (c)), and isothermal compressibility (panel (d)) as functions of $T$ at a few values of $\mu$. We fix $m r_0=1$, $W=0.5\,m$, and $j_{\rm max}=50$, as well as use $n_{\rm max} =2j_{\rm max} +1$.
}
\label{fig:td-T}
\end{figure*}

We show the thermodynamic quantities as functions of $T$ for a few values of $\mu$ in Fig.~\ref{fig:td-T}. To avoid the almost linear growth with the cutoff $j_{\rm max}$ caused by the accumulation of the states with higher values of $j$, we find it convenient to show the thermodynamic quantities normalized by $n_{\rm max} =2j_{\rm max} +1$. The scaling of $\Omega_{V}$ with temperature is linear at large $T$ and reaches a constant value at small $T$. The entropy density rises with $T$ but saturates at larger $T$. The heat capacity is nonmonotonic with a well-pronounced maximum. Finally, the isothermal compressibility diverges at $T\to0$ when the chemical potential is negative. It is worth noting that such a behavior is similar to the thermodynamic properties of systems with a finite number of available states above any given energy, such as a classical 1D nearest-neighbor Ising chain. The flat density of states for the impurity states with $\epsilon_j(V_0) \sim -V_0/2$ further enhances the similarity.

\begin{figure*}[!ht]
\centering
\subfigure[]{\includegraphics[width=0.24\textwidth]{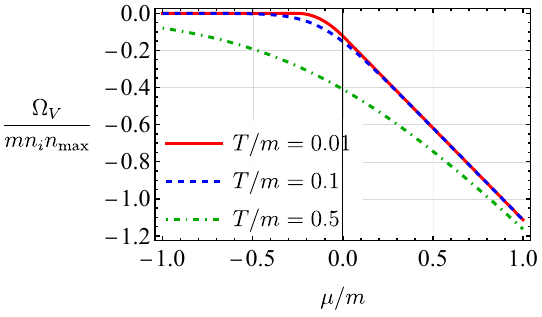}}
\subfigure[]{\includegraphics[width=0.24\textwidth]{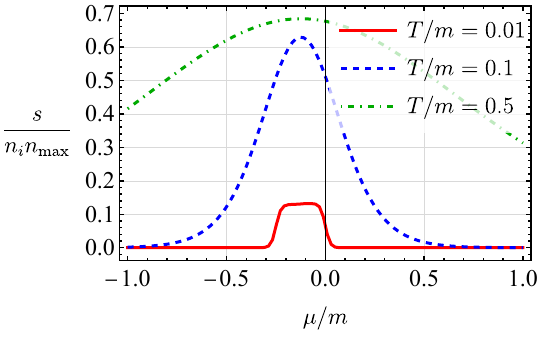}}
\subfigure[]{\includegraphics[width=0.24\textwidth]{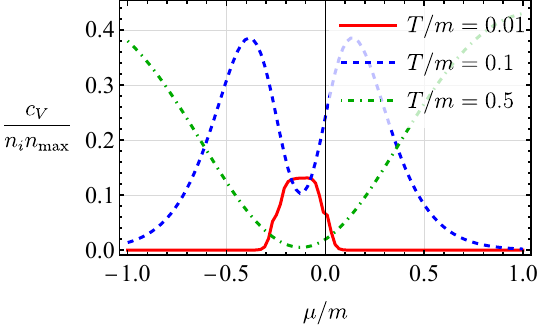}}
\subfigure[]{\includegraphics[width=0.24\textwidth]{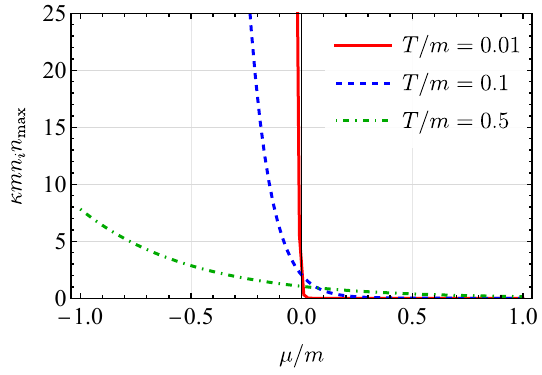}}
\caption{
The thermodynamic potential (panel (a)), entropy density (panel (b)), heat capacity (panel (c)), and isothermal compressibility (panel (d)) as functions of $\mu$ at a few values of $T$. We fix $m r_0=1$, $W=0.5\,m$, and $j_{\rm max}=50$, as well as use $n_{\rm max} =2j_{\rm max} +1$. 
}
\label{fig:td-mu}
\end{figure*}

We display the thermodynamic quantities for the bound states, which decouple from the flat band, as functions of $\mu$ for a few values of $T$ in Fig.~\ref{fig:td-mu}. There is a strong dependence on the sign of $\mu$ because these bound states have $\epsilon<0$. For example, the entropy density and heat capacity are nontrivial only for certain values of $\mu$. The dependence on the chemical potential distinguishes the thermodynamic properties of bound states from those of the 1D Ising model.

\section{Gapless pseudospin-1 model}
\label{sec:gapless}

In the absence of an energy gap, there are no true bound states in the system. Still, impurities result in additional states known as impurity resonances~\cite{Balatsky-Zhu:rev-2006}. The most direct manifestation of resonances is the presence of additional peaks in the LDOS, which is the main focus of this section. We pay special attention to the resonances with $-V_0 <\mbox{Re}\, \epsilon < 0$ since they are related to the bound states that decouple from the flat band.

\subsection{Flat band states}

Let us consider whether there are flat band states in the potential well $V(r)=-V_0\theta(r_0-r)$. In the gapped pseudospin-1 case, an affirmative answer was obtained in Ref.~\cite{Gorbar-Oriekhov:2018}. Let us begin with the three-component system of equations~\cite{Gorbar-Oriekhov:2018}
\begin{eqnarray}
c^{\prime}+\frac{j-1}{r}c+\sqrt{2}\left[V(r)-\epsilon\right] a=0,
\label{system-1}\\
a^{\prime}-\frac{j}{r}a+b^{\prime}+\frac{j}{r}b-\sqrt{2}\left[V(r)-\epsilon\right]c=0,
\label{system-2}\\
c^{\prime}-\frac{j+1}{r}c+\sqrt{2}\left[V(r)-\epsilon\right]b=0.
\label{system-gapless}
\end{eqnarray}
For $r>r_0$ and $\epsilon=0$, Eqs.~(\ref{system-1}) and (\ref{system-gapless}) result in $c_>(r)=0$ with $a_>(r)$ and $b_>(r)$ related by the equation $(a_>(r)+b_>(r))^{\prime}=j(a_>(r)-b_>(r))/r$ leaving one of these components completely arbitrary. For $r<r_0$ and $\epsilon=0$, we find $c_<=C_1rJ_j(V_0r)$, $a_< = [c^{\prime}_<+(j-1)c_</r]/(\sqrt{2}\,V_0)$, and $b_< = [c^{\prime}_<-(j+1)c_</r]/(\sqrt{2}\,V_0)$. Matching conditions
\begin{equation}
c_<(r_0)=c_>(r_0),\quad a_<(r_0)+b_<(r_0)=a_>(r_0)+b_>(r_0)
\label{gapless-matching}
\end{equation}
are satisfied either for 
\begin{align}
&(i)\quad C_1=0,\quad a_>(r_0)+b_>(r_0)=0
\label{solution-1}
\end{align}
or
\begin{align}
&(ii)\qquad J_j(V_0r_0)=0, \nonumber\\
&a_>(r_0)+b_>(r_0)=\sqrt{2}C_1 r_0 \left[J^{\prime}_j(x)-\frac{J_j(x)}{x}\right]_{|_{x=V_0r_0}}.
\label{solution-2}
\end{align}
In the absence of a potential, flat-band states are given by solutions with $a(r)$ and $b(r)$ related by the equation $(a(r)+b(r))^{\prime}=j(a(r)-b(r))/r$ and $c(r)=0$. Therefore, Eq.~\eqref{solution-1} corresponds to flat band states restricted by one additional condition $a_>(r_0)+b_>(r_0)=0$. Solutions described by Eq.~\eqref{solution-2} are realized for particular values of $V_0r_0$ for which $J_j(V_0r_0)=0$. Thus, we conclude that, like in the gapped case, practically all flat band states survive in the gapless pseudospin-1 model with the additional restriction $a_>(r_0)+b_>(r_0)=0$.

\subsection{Resonances}
\label{sec:gapless-res}

To determine resonances for potential well $V(r)=-V_0\theta(r_0-r)$, we consider solutions to
\begin{equation}
f^{\prime\prime}+\frac{f^{\prime}}{r}-\frac{j^2f}{r^2}+(\epsilon-V)^2f+\frac{f^{\prime}V^{\prime}}{\epsilon-V}=0.
\label{equation-component-c-gapless-1}
\end{equation}
This equation is derived from Eqs.~\eqref{system-1}--\eqref{system-gapless} where $c(r) = r f(r)$. Note that, similar to Eqs.~(\ref{equation-component-f-1}) and (\ref{equation-component-c-upper-1}) describing bound states in the gapped case, Eq.~(\ref{equation-component-c-gapless-1}) is singular at the point $x$ where $V(x)=\epsilon$. This point is not a turning point in the gapless case where solutions exist for any $\epsilon$ in the absence of a potential. At $r<r_0$, the solution regular at the origin is $f_<(r)= C_1J_j(\sqrt{(\epsilon+V_0)^2\,}r)$. We find it convenient to search for solutions at $r>r_0$ in the form $f_>(r)=C_3H_j^{(1)}(\sqrt{\epsilon^2} r)+C_4H_j^{(2)}(\sqrt{\epsilon^2} r)$, where $H_j^{(1,2)}(x)$ are the Hankel functions of the first and second kind, respectively. Note that, since we seek resonances where $\epsilon$ is complex, we use $\sqrt{\epsilon^2}$ instead of $|\epsilon|$.

The matching conditions \eqref{gapless-matching} give
\begin{equation}
f_>(r_0)=f_<(r_0),\qquad \frac{f^{\prime}_>(r_0)}{\epsilon}=\frac{f^{\prime}_<(r_0)}{\epsilon+V_0}
\label{matching-gapless}
\end{equation}
and result in the following equations:
\begin{eqnarray}
\label{matching-gapless-H-1}
&&C_1 J_j(\sqrt{(\epsilon+V_0)^2}r_0) \nonumber\\
&&= C_3H_j^{(1)}(\sqrt{\epsilon^2} r_0)+C_4H_j^{(2)}(\sqrt{\epsilon^2} r_0),\\
\label{matching-gapless-H-2}
&&C_1 J_j'(\sqrt{(\epsilon+V_0)^2}r_0) \frac{\sqrt{(\epsilon+V_0)^2}}{V_0+\epsilon} \nonumber\\
&&= C_3 \left( H_j^{(1)}(\sqrt{\epsilon^2} r_0)\right)' \frac{\sqrt{\epsilon^2}}{\epsilon} +C_4 \left(H_j^{(2)}(\sqrt{\epsilon^2} r_0)\right)' \frac{\sqrt{\epsilon^2}}{\epsilon},\nonumber\\
\end{eqnarray}
where the prime denotes the derivative with respect to the argument.

\begin{figure}[!ht]
\centering
\includegraphics[width=0.45\textwidth]{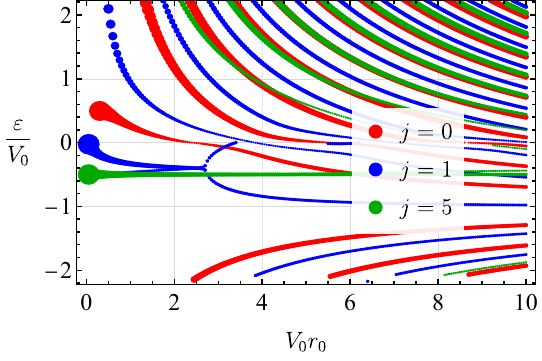}
\caption{
The solutions to Eq.~\eqref{matching-gapless-H-res} for a few values of $j$ as a function of $V_0r_0$. We show the real parts of solutions, see Eq.~\eqref{matching-gapless-H-res-2} for $\Re{\epsilon}>0$ and the text below it for $\Re{\epsilon}<0$. The size of points reflects the magnitude of the imaginary part.
}
\label{fig:gapless-epsilon}
\end{figure}

To find resonances, we focus only on the outgoing solutions at $r>r_0$, therefore, we set $C_4=0$ for $\Re{\epsilon} >0$ and $C_3=0$ for $\Re{\epsilon} <0$. Then Eqs.~\eqref{matching-gapless-H-1} and \eqref{matching-gapless-H-2} are reduced to the following form at $\Re{\epsilon} >0$:
\begin{eqnarray}
\label{matching-gapless-H-res}
&&\frac{\epsilon}{V_0+\epsilon} \frac{\sqrt{(\epsilon+V_0)^2}}{\sqrt{\epsilon^2}}\left(\ln{J_j(\sqrt{(\epsilon+V_0)^2}r_0)}\right)' \nonumber\\
&&= \left(\ln{H_j^{(1)}(\sqrt{\epsilon^2} r_0)}\right)'
\end{eqnarray}
or, alternatively,
\begin{eqnarray}
\label{matching-gapless-H-res-2}
&&\frac{\sqrt{(\epsilon+V_0)^2}}{\epsilon+V_0}\frac{J_{j-1}(\sqrt{(\epsilon+V_0)^2}r_0)}{J_j(\sqrt{(\epsilon+V_0)^2}r_0)} - \frac{\sqrt{\epsilon^2}}{\epsilon} \frac{H_{j-1}^{(1)}(\sqrt{\epsilon^2}r_0)}{H_j^{(1)}(\sqrt{\epsilon^2} r_0)} \nonumber\\
&&+ \frac{jV_0}{\epsilon r_0(V_0+\epsilon)}=0.
\end{eqnarray} 
For $\Re{\epsilon} <0$, one should replace $H_{j}^{(1)}(\sqrt{\epsilon^2} r_0)$ with $H_{j}^{(2)}(\sqrt{\epsilon^2}  r_0)$ and use the branch $\sqrt{\epsilon^2} =-\epsilon$.
This characteristic equation has complex roots $\epsilon = \epsilon_r -i\Gamma/2$ with $\Gamma>0$ defining the width of the resonances. The numerical solutions to Eq.~\eqref{matching-gapless-H-res} are shown in Fig.~\ref{fig:gapless-epsilon}. As one can see, there are several roots even at $j=0$. The energies tend to decrease with $V_0r_0$, leading to denser energy levels with a smaller imaginary part. As with the bound states shown in Fig.~\ref{fig:levels}, there are resonances with $\Re{\epsilon} \approx -V_0/2$ for $j>0$. Unlike other resonances, they are characterized by a much smaller imaginary part and, hence, are more reminiscent of their bound-state counterparts studied in Sec.~\ref{sec:gapped-wf}.

Resonances can also be found by considering the scattering phase $\delta_j$.
To obtain the scattering phases, one searches for a solution in the region $r>r_0$ as $f_>(r)=C_5 \cos{(\delta_j)} \, J_j(|\epsilon| r) - C_5 \sin{(\delta_j)} \,Y_j(|\epsilon| r)$. This solution has correct asymptotics at $r\to \infty$. Then, matching the wave functions at $r=r_0$, we obtain
\begin{eqnarray}
\label{matching-gapless-delta-1}
C_1 J_j(|V_0+\epsilon|r_0) &=& C_5\cos{(\delta_j)} \, J_j(|\epsilon| r_0) \nonumber\\
&-&C_5\sin{(\delta_j)} \, Y_j(|\epsilon| r_0),
\end{eqnarray}
\begin{eqnarray}
\label{matching-gapless-delta-2}
&&C_1 J_j'(|V_0+\epsilon|r_0) \sign{V_0+\epsilon} \nonumber\\
&&= C_5\cos{(\delta_j)} \, J_j'(|\epsilon| r_0) \sign{\epsilon} \nonumber\\
&&-C_5 \sin{(\delta_j)} \, Y_j'(|\epsilon| r_0) \sign{\epsilon}.
\end{eqnarray}
These equations lead to the following characteristic equation for the scattering phases $\delta$:
\begin{eqnarray}
\label{gapless-delta-sol}
&&\tan{\delta_j(\epsilon)} = \nonumber\\
&&\frac{J_j'(|\epsilon| r_0) -J_j(|\epsilon| r_0) \sign{\frac{\epsilon}{\epsilon +V_0}}\left(\ln{J_j(|V_0+\epsilon|r_0)}\right)' }{Y_j'(|\epsilon| r_0) -Y_j(|\epsilon| r_0) \sign{\frac{\epsilon}{\epsilon +V_0}} \left(\ln{J_j(|V_0+\epsilon|r_0)}\right)'}.\nonumber\\
\end{eqnarray}

Let us proceed now to the calculation of the local density of states.

\subsection{LDOS}
\label{sec:green}

Impurity resonances are manifested and have a pronounced effect on the local DOS (LDOS)~\cite{Balatsky-Zhu:rev-2006}. The LDOS of dispersive states with $\epsilon \ne 0$ is defined as
\begin{equation}
\label{green-ldos-def}
\nu(\epsilon, \mathbf{r}) = \sum_j\psi_{\epsilon,j}^{\dag}(\mathbf{r})\psi_{\epsilon,j}(\mathbf{r}).
\end{equation}
The same expression can be obtained from a Green's function, see, e.g., Ref.~\cite{Balatsky-Zhu:rev-2006}. Therefore, the LDOS requires one to know only the properly normalized wave functions in the inner and outer regions of the potential well. Under ``properly normalized wave functions", we understand the wave functions that have correct asymptotics at $r\to \infty$, which is fixed by the relation
\begin{equation}
\label{green-psi2-norm}
\int_0^{2\pi} d\varphi \int_0^{\infty}r\,dr \,\psi_{\epsilon,j}^{\dag}(\mathbf{r})\psi_{\epsilon',j'}(\mathbf{r}) = \df{\epsilon-\epsilon'} \delta_{jj'}.
\end{equation}

To determine the normalization constant, we use the following expression for the second component of the wave function $c(r)$ at $r>r_0$:
\begin{eqnarray}
\label{green-wf-asym-rgr0}
&&C_5 r\cos{(\delta_j)} \, J_j(|\epsilon| r) -C_5 r\sin{(\delta_j)} \, Y_j(|\epsilon| r) \nonumber\\
&&\simeq C_5 r \sqrt{\frac{2}{\pi |\epsilon|r}} \cos{\left(|\epsilon| r -\frac{\pi j}{2} -\frac{\pi}{4} +\delta_j\right)},
\end{eqnarray}
where $\delta_j$ is the scattering phase defined in Eq.~\eqref{gapless-delta-sol}. The first and third components can be found from Eqs.~\eqref{system-1} and \eqref{system-gapless}, respectively. Calculating $\psi_{\epsilon,j}^{\dag}(\mathbf{r})\psi_{\epsilon',j'}(\mathbf{r})$, using the asymptotic form for the Bessel functions, and neglecting all terms that do not produce $\delta$-functions, we arrive at
\begin{equation}
\label{green-C5}
|C_5| = \sqrt{\frac{|\epsilon|}{4\pi}}.
\end{equation}
Coefficient $C_1$ is fixed from Eq.~\eqref{matching-gapless-delta-1}.

Thus, the wave functions at $r<r_0$ and $r>r_0$ are given by
\begin{widetext}
\begin{eqnarray}
\label{green-wf-2-rle0}
\psi_{\epsilon,j;2}(r<r_0) &=& i e^{ij\varphi} \sqrt{\frac{|\epsilon|}{4\pi}}
\frac{\cos{(\delta_j)} \, J_j(|\epsilon| r_0) -\sin{(\delta_j)} \, Y_j(|\epsilon| r_0)}{J_j(|V_0+\epsilon|r_0)} J_j(|V_0+\epsilon|r),\\ 
\label{green-wf-2-rge0}
\psi_{\epsilon,j;2}(r>r_0) &=& i e^{ij\varphi} \sqrt{\frac{|\epsilon|}{4\pi}}
\left[\cos{(\delta_j)} \, J_j(|\epsilon| r) -\sin{(\delta_j)} \, Y_j(|\epsilon| r) \right].
\end{eqnarray}
\end{widetext}
The rest of the components, $\psi_{\epsilon,j;1}$ and $\psi_{\epsilon,j;3}$, follow from the following equations:
\begin{widetext}
\begin{eqnarray}
\label{green-wf-1-rle0}
\psi_{\epsilon,j;1}(r<r_0) &=& -e^{-i\varphi} \frac{1}{r \sqrt{2}(V_0+\epsilon)} \left[ \partial_r\left(r \psi_{\epsilon,j;2}(r<r_0)\right) +(j-1)\psi_{\epsilon,j;2}(r<r_0)\right],\\
\label{green-wf-3-rle0}
\psi_{\epsilon,j;3}(r<r_0) &=& -e^{i\varphi} \frac{1}{r \sqrt{2}(V_0+\epsilon)} \left[\partial_r\left(r \psi_{\epsilon,j;2}(r<r_0)\right) -(j+1) \psi_{\epsilon,j;2}(r<r_0)\right],\\
\label{green-wf-1-rge0}
\psi_{\epsilon,j;1}(r>r_0) &=& -e^{-i\varphi}\frac{1}{r \sqrt{2}\epsilon} \left[\partial_r\left(r \psi_{\epsilon,j;2}(r>r_0)\right) +(j-1) \psi_{\epsilon,j;2}(r>r_0)\right],\\
\label{green-wf-3-rge0}
\psi_{\epsilon,j;3}(r>r_0) &=& -e^{i\varphi} \frac{1}{r \sqrt{2}\epsilon} \left[\partial_r\left(r \psi_{\epsilon,j;2}(r>r_0)\right) -(j+1) \psi_{\epsilon,j;2}(r>r_0) \right].
\end{eqnarray}
\end{widetext}
The scattering phases $\delta_j(\epsilon)$ follow from Eq.~\eqref{gapless-delta-sol}.

So far, we have not taken into account the zero-energy solution. According to the discussion after Eq.~\eqref{system-gapless}, the normalizable solution corresponding to $\epsilon=0$ can be taken in the following form:
\begin{eqnarray}
\label{green-wf-0-1}
&&\psi^T(r>r_0) = \frac{e^{ij\phi}}{r} \left(a_>e^{-i\phi}, 0, b_{>}e^{i\varphi}\right),\nonumber \\
&&\psi^T(r<r_0) = \frac{e^{ij\phi}}{r} \nonumber\\
&&\times\left(\frac{c^{\prime}_<+(j-1)c_</r}{\sqrt{2}\,V_0}e^{-i\varphi}, ic_{<}, \frac{c^{\prime}_<-(j+1)c_</r}{\sqrt{2}\,V_0}e^{i\varphi}\right), \nonumber\\
\end{eqnarray}
where $b_{>}(r)$ is an arbitrary function, $a_{>}(r)$ is defined by the equation $a^{\prime}_>=ja_>/r-b^{\prime}_>-jb_>/r$, and $c_{<}(r) = C_1 r J_j(r V_0)$. The solution to the equation for $a_>$ contains a solution to a homogeneous equation plus a solution to an inhomogeneous equation, i.e., $a_>(r)=r^jC_a-r^j \int^r dt (b^{\prime}_>(t)+jb_>(t)/t)/t^j$, where $C_a$ is a constant. The matching condition in Eq.~\eqref{solution-1} gives a trivial solution inside the circular potential well. Since the flat band states are characterised by $c(r)=0$ in the absence of a potential, we conclude that practically all these solutions survive in the presence of a circular potential well for any $V_0$ with an additional condition $a_>(r_0)+b_>(r_0)=0$. This condition can be easily resolved by fixing the constant $C_a$ and leaving the function $b_>(r)$ completely arbitrary. On the other hand, the matching condition (\ref{solution-2}) means that nontrivial solutions exist only for certain values of $V_0$. 

\begin{figure*}[!ht]
\centering
\includegraphics[width=0.99\textwidth]{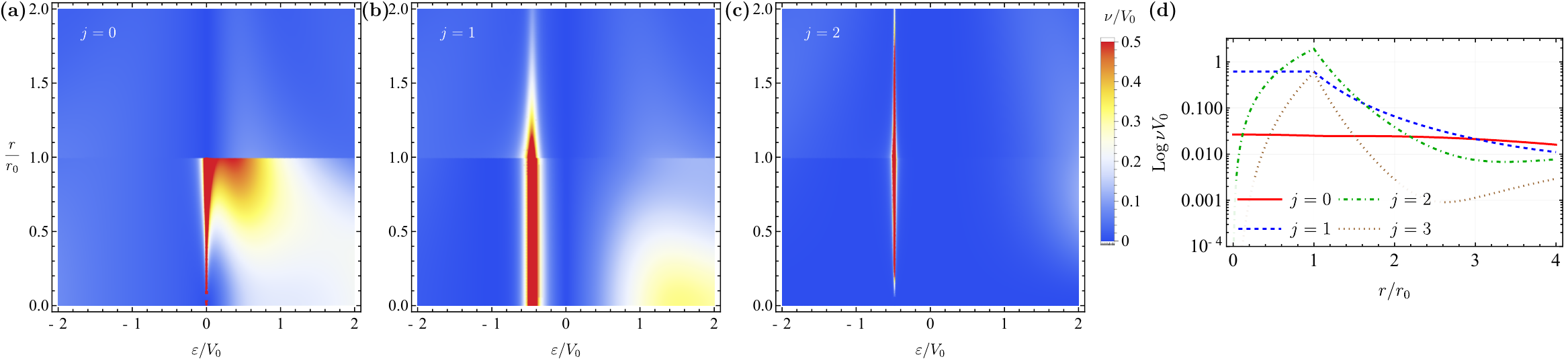}
\caption{
Panels (a)--(c): The LDOS as a function of energy $\epsilon$ for the states with $j=0,1,2$. Panel (d): The LDOS at $\epsilon=-V_0/2$ as a function of $r/r_0$ on a log scale for the states with $j=0,1,2$. In all panels, we fixed $V_0r_0=1$ and used only the dispersive states.
}
\label{fig:green-ldos-2d}
\end{figure*}

The LDOS $\nu(\epsilon,\mathbf{r})$ as a function of energy $\epsilon$ and radial coordinate $r$ is shown in Figs.~\ref{fig:green-ldos-2d}(a)--\ref{fig:green-ldos-2d}(c). As one can see, only the resonances with $\Re \epsilon <0$ are pronounced outside the potential well and all resonances contribute inside the potential well. Narrow straight maxima of the LDOS for $j>0$ correspond to resonances with $\Re \epsilon \approx -V_0/2$, see also Fig.~\ref{fig:gapless-epsilon}. 

To contrast the localization properties of bound states and resonances, we also plot the LDOS for resonances at fixed $\epsilon=-V_0/2$ in Fig.~\ref{fig:green-ldos-2d}(d). Similar to the bound states that decouple from the flat band, see Fig.~\ref{fig:psi2}(a), resonances with $j>1$ and $\epsilon \approx -V_0/2$ tend to be localized at the boundary of the potential well, in agreement with the singularity of the term with $f^{\prime}V^{\prime}/(\epsilon-V)$ in the effective one-component equation, and have discontinuous LDOS at $r=r_0$. The resonance with $j=0$ shows a distinct behavior.

\section{Summary and discussion}
\label{sec:conclusion}

We investigated the formation and properties of bound states and resonances in pseudospin-1 systems. The presence of a flat band allows for a different set of properties compared to the impurity states in Dirac materials. In the case of gapped pseudospin-1 fermions, an impurity potential gives rise to two different types of bound states: those that decouple from the flat band and those that decouple from the dispersive band, see Fig.~\ref{fig:levels}(b). 

The bound states in the vicinity of the flat band in a circular potential well of depth $V_0$ for gapped pseudospin-1 fermions are strongly localized at the edge of the potential well $r_0$ and, thus, resemble ``whispering-gallery" modes circling the well, see Fig.~\ref{fig:psi2}(a). Another distinct property of these bound states is related to their dependence on the total angular momentum $j$. Unlike gapped Dirac systems, where there are no bound states with the angular momentum exceeding a certain critical value, bound states for pseudospin-1 systems exist for \emph{any}~\footnote{In realistic lattice systems there is a cutoff for the angular momentum.} value of $j$. Furthermore, their energy \emph{decreases} for higher angular momenta approaching $-V_0/2$, while their localization at $r_0$ \emph{increases} with $j$. The mathematical reason for these unusual characteristics is the singular character of the interaction term proportional to $f^{\prime}V^{\prime}/(\epsilon-V)$ in the effective one-component equation for pseudospin-1 quasiparticles at the turning point in the vicinity of the flat band. This is a unique feature of multicomponent pseudospin-1 fermions rather than a consequence of the presence of a flat band, as is supported by the analysis of bound states in gapped graphene in a magnetic field, see Appendix~\ref{sec:tmd}. Although the magnetic field realizes dispersionless Landau levels, bound states induced by a potential well are not localized at its edge in agreement with the regular character of the $g^{\prime}V^{\prime}/(\epsilon-V -\Delta)$ interaction term in the effective one-component equation for electron quasiparticles in gapped graphene.

As to bound states near the upper band for gapped pseudospin-1 fermions, we found that although the centrifugal barrier squeezes them to the edge as $j$ increases~\footnote{The squeezing of the bound states near the upper band for gapped pseudospin-1 fermions is also related to the singular term with $f^{\prime}V^{\prime}/(\epsilon -m -V)$ in the corresponding effective one-component equation.}, see Fig.~\ref{fig:psi2}(b), these states exist only up to the critical angular momentum $j_c$, and their energies grow with $|j|$ while approaching the band continuum as $|j|$ increases. This property resembles that for bound states in gapped graphene and in the case of the conventional Schr\"{o}dinger equation.

Assuming the Anderson model of disorder, we also studied the contribution of the bound states that split off the flat band to the thermodynamic properties, see Sec.~\ref{sec:thermodynamics}. We found that for the chemical potential inside the gap, the temperature dependence of the thermodynamic potential, entropy density, heat capacity, and isothermal compressibility is similar to that for systems with a finite number of available states above any given energy, such as a classical 1D nearest-neighbor Ising chain. The entropy density monotonically increases and then saturates with temperature, while the heat capacity attains a single maximum as a function of temperature, see Fig.~\ref{fig:td-T}.

When the gap closes, the bound states do not disappear but transform into resonances. These states acquire a finite imaginary part and show similar dependence on the angular momentum as in the gapped case, see Fig.~\ref{fig:gapless-epsilon}. The LDOS demonstrates resonance-induced enhancement either inside the well or near its boundary, see Fig.~\ref{fig:green-ldos-2d}. The peaks of the LDOS at the well edge are observed for the states with $j>1$ and energies $\Re \epsilon \approx -V_0/2$. This agrees with the singularity of the term with $f^{\prime}V^{\prime}/(\epsilon-V)$ in the effective one-component equation describing these resonances.

Our results address the fundamental properties of the impurity states in the pseudospin-1 systems. Furthermore, we provide potentially verifiable signatures of the impurity states in thermodynamic quantities of pseudospin-1 fermions and the LDOS.

\section{Acknowledgments}

The work of E.V.G. acknowledges support from the National Research Foundation of Ukraine grant (2023.0097) “Electronic and transport properties of Dirac materials and Josephson junctions”. 
The authors thank V.P.~Gusynin and D.O.~Oriekhov for useful discussions.

\appendix

\vspace{5mm}

\begin{widetext}

\section{Bessel-function identities and asymptotic expansions}
\label{sec:App-Bessel}

For completeness, we provide a few useful relations for the Bessel functions used in our work, see Ref.~\cite{Olver}. We use the following recurrence relations:
\begin{equation}
I_{j-1}(x)-I_{j+1}(x) =\frac{2j}{x}I_{j}(x),
\quad I_{j-1}(x)+I_{j+1}(x) = 2I^{\prime}_{j}(x),
\label{I-function}
\end{equation}
\begin{equation}
K_{j-1}(x)-K_{j+1}(x) =-\frac{2j}{x}K_{j}(x), \quad K_{j-1}(x)+K_{j+1}(x) =-2 K^{\prime}_{j}(x).
\label{K-function}
\end{equation}

The asymptotic behavior at small $x$ differs for $j=0,1$ and $j>1$. For $j>1$, we have
\begin{equation}
I_j(x) \sim \left(\frac{x}{2}\right)^j\frac{1}{j!}\left[1+\frac{x^2}{4(j+1)}\right],\quad K_j(x) \sim \frac{2^{j-1}(j-1)!}{x^j}\left[1-\frac{x^2}{4(j-1)}\right]
\label{asymptotics-small}
\end{equation}
and, for $j=0$ and $j=1$, the small $x$ asymptotics are
\begin{equation}
I_0(x)\sim 1+\frac{x^2}{4},\quad I_1(x)\sim \frac{x}{2}+\frac{x^3}{16},\quad K_0(x)\sim -\gamma -\ln\left(\frac{x}{2}\right)+\frac{x^2}{4}\left(1-\gamma-\ln\frac{x}{2}\right),\quad K_1(x)\sim \frac{1}{x}+\frac{x}{2}\left(\ln\frac{x}{2}+\gamma-\frac{1}{2}\right),
\label{energy-asym}
\end{equation}
where $\gamma$ is Euler's constant.

At large $x$, the asymptotics are
\begin{align}
I_{j}(x) \sim \frac{e^x}{\sqrt{2\pi x}}\left(1-\frac{4j^2-1}{8x} +\frac{(4j^2-1)(4j^2-9)}{128 x^2}\right),\quad K_{j}(x) \sim \sqrt{\frac{\pi}{2x}}e^{-x}\left(1+\frac{4j^2-1}{8x} +\frac{(4j^2-1)(4j^2-9)}{128 x^2}\right).
\label{function-asymptotics}
\end{align}

For $j \to +\infty$, the asymptotics of functions $I_j(x)$ and $K_j(x)$ are given by
\begin{equation}
\label{app-asymptotics-large-j}
I_j(x)\sim \frac{1}{\sqrt{2\pi j}}\left(\frac{ex}{2j}\right)^j\left[1-\frac{1}{12j}+\frac{x^2}{4(j+1)}\right],\quad 
K_j(x)\sim \sqrt{\frac{\pi}{2 j}} \left(\frac{ex}{2j}\right)^{-j}\left[1+\frac{1}{12j}-\frac{x^2}{4(j-1)}\right].
\end{equation}

\section{Bound state energies for regularized potential}
\label{sec:App-1}

Approximating $(\epsilon-V)^2-m^2$ with $-m^2$ and neglecting $(\epsilon-V)^2$ compared to $m^2$ for states in the vicinity of the flat band, Eq.~(\ref{equation-component-f-1}) in terms of the dimensionless variable $x=mr$ takes the form
\begin{equation}
-\left(f^{\prime\prime}+\frac{f^{\prime}}{x}-\frac{j^2f}{x^2}-f-\frac{2jfV^{\prime}}{mx}\right)(\epsilon-V)+f^{\prime}V^{\prime}=0.
\label{equation-component-c-2}
\end{equation}
One of the key characteristics of this equation is that the first derivative of $f$ equals zero at the turning point $x_{t}$ defined by the equation $\epsilon=V(x_{t})$, i.e., $
f^{\prime}(x_{t})=0$. Mathematically, a point where the coefficient of the highest-order derivative becomes zero is a singular point of the corresponding differential equation. Since the wave function of a bound state should decrease in the classically forbidden region where $\epsilon <V(r)$, this suggests that $f$ should attain a maximum at $x_t$. For the potential well with a sharp edge, the turning point is obviously given by $x_0=mr_0$, and Fig.~\ref{fig:psi2}(a) confirms that the wave function indeed reaches the maximum at $x_0$.

Thus, it is instructive to find a solution to the above equation in the vicinity of a turning point, where Eq.~(\ref{equation-component-f-1}) takes the form
\begin{equation}
f^{\prime\prime} -f^{\prime}\frac{V^{\prime}}{\epsilon-V}=0.
\label{turning-point}
\end{equation}
Its solution is given by 
\begin{equation}
f^{\prime} = \frac{A}{\epsilon-V}
\end{equation}
and $f^{\prime}$ diverges at the turning point. 

To find an analytical solution to Eq.~(\ref{equation-component-c-2}),
we approximate potential (\ref{regularized}) for large $\lambda$ as follows:
\begin{equation}
V_{\lambda}(x) \approx -V_0\theta\left(x_1-x\right)-\frac{V_0}{2}\left[1-\frac{\lambda(x-x_0)}{x_0}\right]\,\theta\left(x_2-x\right)\theta\left(x-x_1\right),\quad x_1=x_0\left(1-\frac{1}{\lambda}\right), \quad x_2=x_0\left(1+\frac{1}{\lambda}\right).
\label{potential-approximate}
\end{equation}

In region I, where $0<x<x_1$, potential $V_{\lambda}(x)$ is constant and equals $-V_0$. Then Eq.~(\ref{equation-component-c-2}) takes the form of the modified Bessel equation whose normalizable solution at the origin is given by $f_I(x)=C_1I_j(x)$. Similarly, we find the normalizable solution $f_{III}(x)=C_3K_j(x)$ in region III, where $x>x_2$.

In region II, where $x_1<x<x_2$, Eq.~(\ref{equation-component-c-2}) acquires the form
\begin{equation}
-\left(f^{\prime\prime}+\frac{f^{\prime}}{x}-\frac{j^2f}{x^2}-f-\frac{2jfA}{mx}\right)\left[\epsilon+\frac{V_0}{2}-A(x-x_0)\right]+f^{\prime}A=0,\qquad A=\frac{\lambda V_0}{2x_0}.
\label{equation-region-2}
\end{equation}
Since this region is very small for $\lambda \gg 1$, it is useful to make the change of variable $x-x_0=yx_0/\lambda$, where $y$ takes values in the interval $[-1,1]$. Then, keeping terms of order $1/\lambda$ and neglecting higher order terms, Eq.~(\ref{equation-region-2}) reads
\begin{equation}
-\left[f^{\prime\prime}+\frac{f^{\prime}}{\lambda} -\frac{jfV_0}{m\lambda}\right]\left(\epsilon+\frac{V_0}{2}-\frac{V_0y}{2}\right)+\frac{f^{\prime}V_0}{2}=0,
\label{equation-region-3}
\end{equation}
where we approximated $x^2$ and $x$ in denominator by $x^2_0$ and $x_0$, respectively. In terms of variable $z=-(\frac{2\epsilon}{V_0}+1-y)$ which takes values in the interval $[-\frac{2\epsilon}{V_0}-2,-\frac{2\epsilon}{V_0}]$, Eq.~(\ref{equation-region-3}) acquires the form
\begin{equation}
f^{\prime\prime}+\left(\frac{1}{z}+\frac{1}{\lambda}\right)f^{\prime}
-\frac{jfV_0}{m\lambda}=0.
\label{equation-region-4}
\end{equation}
Its general solution is expressed in terms of the Whittaker functions~\cite{Olver}
\begin{equation}
f(z)=z^{-1/2} e^{-z/(2\lambda)} \left[C_0M_{-\kappa,0}\left(qz\right)+C_2W_{-\kappa,0}\left(qz\right)\right], 
\label{solution-second}
\end{equation}
where $\kappa=1/\sqrt{4+16j\lambda V_0/m}$ and $q=1/(2\lambda \kappa)$. Using the expansion~\cite{Olver}
\begin{equation}
M_{\kappa,0}(x)=x^{1/2}-\kappa x^{3/2}+ \frac{1+4\kappa^2}{16} x^{5/2} +\ldots,
\end{equation}
\begin{equation}
W_{\kappa,0}(x)=-\frac{x^{1/2}}{\Gamma(\frac{1}{2}-\kappa)}L_{\kappa}(x) +\frac{\kappa x^{3/2}}{\Gamma(\frac{1}{2}-\kappa)}\left[L_{\kappa}(x) -2\right]
-\frac{x^{5/2}}{16\,\Gamma(\frac{1}{2}-\kappa)} \left[(1+4\kappa^2)L_{\kappa}(x) -(1+12\kappa^2)\right]
+ \ldots
\label{W-function}
\end{equation}
at $x \to 0$, where $\gamma$ is the Euler constant and $L_{\kappa}(x) = \ln x+\psi\left(\frac{1}{2}-\kappa\right)+2\gamma$, we obtain
\begin{eqnarray}
f(z) &=& \frac{e^{-z/(2\lambda)}}{\sqrt{2\kappa\lambda}} \Bigg\{ C_0\left( 1+\frac{z}{2\lambda} +\frac{jV_0z^2}{4m\lambda} \right) \nonumber\\ 
&-& \frac{C_2}{\Gamma(\frac{1}{2}+\kappa)} \Bigg[ \left( 1+\frac{z}{2\lambda} +\frac{jV_0z^2}{4m\lambda} \right) \left( \ln\frac{z}{2\kappa\lambda} +\psi\left(\frac{1}{2}+\kappa\right) +2\gamma \right) 
-\frac{z}{\lambda} -\frac{jV_0z^2}{4m\lambda} \Bigg] \Bigg\}.
\label{solution-all}
\end{eqnarray}

The Whittaker function $W_{-\kappa,0}(x)$ has a branch point at $x=0$, which is reflected by the presence of logarithm terms in the expansion (\ref{W-function}). Clearly, real solution for $x>0$ is $\ln x$ and $\ln(-x)$ for $x<0$. To account for these real branches, we replace $\ln \frac{z}{2\kappa\lambda} \to \frac{1}{2}\ln \left(\frac{z}{2\kappa\lambda}\right)^2$. Then, in the linear order in $1/\lambda$, we have
\begin{equation}
f_{II}(z) \approx \tilde{C}_0 \left(1 + \frac{jV_0z^2}{4m\lambda}\right) +\tilde{C}_2\left[\left(1 + \frac{jV_0z^2}{4m\lambda}\right)  \frac{1}{2}\ln z^2 -\frac{z}{\lambda} -\frac{jV_0z^2}{4m\lambda}\right].
\label{solution-second-1}
\end{equation}

By matching solutions $f_I$, $f_{II}$, and $f_{III}$, as well as their derivatives at $x_1$ and $x_2$, we obtain the following characteristic equation:
\begin{equation}
\label{energy-char-eq-0}
x_0^2\left(A_1B_2-A_2B_1\right)\ell_I\ell_K +x_0\lambda
\left[ \left(A_2'B_1-A_1B_2'\right)\ell_I +\left(A_2B_1'-A_1'B_2\right)\ell_K \right] +\lambda^2\left(A_1'B_2'-A_2'B_1'\right)=0,
\end{equation}
where 
\begin{eqnarray}
\ell_I &=& \left(\ln I_j(x_1)\right)', \quad
\ell_K=\left(\ln K_j(x_2)\right)', \nonumber\\
A(z) &=& 1+\frac{bz^2}{4\lambda}, \quad
B(z)= \left(1+\frac{bz^2}{4\lambda}\right) \frac{1}{2}\ln z^2 -\frac{z}{\lambda} -\frac{bz^2}{4\lambda}
\end{eqnarray}
with $u=\epsilon/V_0$, $b = jV_0/m$, $z_1=-2(u+1)$, $z_2=-2u$.
The boundary quantities are defined by
\begin{equation}
A_i=A(z_i), \quad A_i'=A'(z_i), \quad B_i=B(z_i), \quad B_i'=B'(z_i).
\end{equation}
with $i=1,2$. Dividing Eq.~\eqref{energy-char-eq-0} by $A_1 A_2$, gathering the terms at $\ell_I -\lambda A_1'/(x_0 A_1)$ and $\ell_K -\lambda A_2'/(x_0 A_2)$, and expanding up to the first order in $1/\lambda$, we derive
\begin{equation}
\label{energy-char-eq}
-\ln\left(1+\frac{V_0}{\epsilon}\right)^2 +\frac{2}{\lambda}
\left[ \frac{jV_0}{m} \left(\frac{2\epsilon}{V_0}+1\right)-2 \right] 
= \frac{\left[\frac{1}{\epsilon/V_0+1} +\frac{2}{\lambda} -\frac{2jV_0}{m\lambda} \left(\frac{\epsilon}{V_0}+1\right) \right]\frac{\lambda}{x_0}}{\left(\ln I_j(x_1)\right)' +\frac{jV_0}{mx_0} \left(\frac{\epsilon}{V_0}+1\right)} - \frac{\left[\frac{1}{\epsilon/V_0} +\frac{2}{\lambda} -\frac{2jV_0}{m\lambda} \frac{\epsilon}{V_0} \right]\frac{\lambda}{x_0} }{\left(\ln K_j(x_2)\right)' +\frac{jV_0}{mx_0} \frac{\epsilon}{V_0}}.
\end{equation}

It is straightforward to find an analytical solution to this equation for $x_0=mr_0 \ll 1$ using asymptotics in Eq.~(\ref{asymptotics-small}). Assuming $j>1$ and expanding Eq.~\eqref{energy-char-eq} to the leading nontrivial order in $x^2_0$, we obtain
\begin{eqnarray}
&&-\ln\left(1+\frac{V_0}{\epsilon}\right)^2 +\frac{2}{\lambda} \left[\frac{jV_0}{m} \left(\frac{2\epsilon}{V_0}+1\right)-2 \right] =\frac{(\lambda-1)\left[\frac{1}{\epsilon/V_0+1} +\frac{2}{\lambda} -\frac{2jV_0}{m\lambda} \left(\frac{\epsilon}{V_0}+1\right) \right]}{j+\frac{jV_0}{m} \left(1-\frac{1}{\lambda}\right) \left(\frac{\epsilon}{V_0}+1\right)} \nonumber\\
&&\times \left[1- \frac{x_0^2\left(1-\frac{1}{\lambda}\right)^2}{2(j+1)\left[j+\frac{jV_0}{m} \left(1-\frac{1}{\lambda}\right) \left(\frac{\epsilon}{V_0}+1\right) \right]}\right] +\frac{(\lambda+1)\left[\frac{1}{\epsilon/V_0} +\frac{2}{\lambda} -\frac{2jV_0}{m\lambda} \frac{\epsilon}{V_0} \right]}{j-\frac{jV_0}{m} \left(1+\frac{1}{\lambda}\right) \frac{\epsilon}{V_0}} \left[1- \frac{x_0^2\left(1+\frac{1}{\lambda}\right)^2}{2(j-1)\left[j-\frac{jV_0}{m} \left(1+\frac{1}{\lambda}\right) \frac{\epsilon}{V_0} \right]} \right]. \nonumber\\
\label{characteristic-equation}
\end{eqnarray}

For $x^2_0 \ll 1$ and $\lambda \to \infty$, this equation has the solution $\epsilon \approx -V_0/2$. Then, using $\epsilon=-V_0\left(1/2 - a_1/\lambda -a_2 x_0^2\right)$, we obtain
\begin{eqnarray}
\label{app-energy-sol-a-x0}
\epsilon &\approx& -\frac{V_0}{2} \left[1 - \frac{b}{2(j+b) \lambda} -\frac{x_0^2}{2(j^2-1)(j+b)}\right]
= -\frac{V_0}{2} \left[ 1-\frac{V_0}{2(m+V_0)\lambda} -\frac{mx_0^2} {2j(j^2-1)(m+V_0)} \right] \nonumber\\
&\stackrel{|V_0|\ll m}{\approx}& -\frac{V_0}{2} \left[1 -\frac{V_0}{2m\lambda} -\frac{x_0^2} {2j(j^2-1)} \right].
\end{eqnarray}

In agreement with the results obtained in Sec.~\ref{sec:energy-levels}, the bound-state energy decreases with $j$ at large $j \lesssim \lambda$. Note that within our approximations, the value of the bound-state energy at $x_0=0$ remains the same as for the sharp potential, i.e., $-V_0/2$. Corrections may appear if the condition $|V_0| \ll m$ is relaxed or the terms $\mathcal{O}(1/\lambda^2)$ are fully retained.

\section{Impurity states in gapped graphene in a magnetic field}
\label{sec:tmd}

To compare the properties of bound states for pseudospin-1 and pseudospin-1/2 fermions, let us calculate the spectrum of the bound states in a gapped graphene subject to a central potential $V(r)$ and an external magnetic field. The presence of the magnetic field allows us to realize dispersionless Landau levels when $V(r)=0$ and, therefore, compare the properties of the bound states related to flat bands.

We use the following Hamiltonian for electron quasiparticles in the valley $K$:
\begin{equation}
\label{tmd-H}
H = v_F \bm{\sigma}\cdot \left(\hat{\mathbf{p}}+\frac{e}{c}\mathbf{A}(\mathbf{r})\right) + \Delta\sigma_z +V(r)
= \begin{pmatrix}
\Delta +V(r) & -iv_F e^{-i \varphi} \left(\partial_r -\frac{i}{r} \partial_{\varphi}+\frac{eBr}{2c}\right) \\
-iv_F e^{i \varphi} \left(\partial_r +\frac{i}{r} \partial_{\varphi}-\frac{eBr}{2c}\right) & -\Delta +V(r)
 \end{pmatrix},
\end{equation}
where $\Delta$ is a gap, $\hat{\mathbf{p}}$ is the momentum operator, and $\mathbf{A}(\mathbf{r})=B(-y,x)/2$ is the vector potential which describes a constant magnetic field in the symmetric gauge. We seek the solution in the form
\begin{equation}
\label{tmd-psi}
\Psi(\mathbf{r}) = \frac{1}{r}
\begin{pmatrix}
a(r) e^{i(j-1)\varphi}\\
i b(r) e^{ij\varphi}
\end{pmatrix}
\end{equation}
and obtain the following equation for $f(r)=b(r)/\sqrt{r}$:
\begin{equation}
-f^{\prime\prime}-\frac{f^{\prime}V^{\prime}}{\epsilon-V-\Delta}+\left[\frac{j^2-1/4}{r^2}-\frac{(j-\frac{1}{2}+\frac{eBr^2}{2c})V^{\prime}}{r(\epsilon-V-\Delta)}+\frac{eB(j-1)}{c}+\frac{e^2B^2r^2}{4c^2}\right]f=\frac{(\epsilon-V)^2-\Delta^2}{v^2_F}f.
\label{equation-generic-1}
\end{equation}
The key difference of this equation compared to Eq.~(\ref{equation-component-f-1}) is the presence of $\Delta$ in the denominators of interaction terms proportional to $V^{\prime}$.

In the symmetric gauge, using the lowest Landau level (LLL) functions $\Psi^{LLL}_{-j}$ with the energy $\epsilon=-\Delta$ whose components are given by \cite{Gusynin}
\begin{equation}
\label{app-ab}
a_{-j}(r)=0,\quad b_{-j}(r)=\frac{r(-1)^{j}}{il_B\sqrt{2\pi j!}}\left(\frac{r^2}{2l^2_B}\right)^{j/2}e^{-r^2/(4l^2_B)},\qquad j=0,1,2, \ldots,
\end{equation}
where $l_B=\sqrt{c/|eB|}$ is the magnetic length, we easily find the following perturbative correction to the LLL energy due to a circular potential well $V(r)=-V_0\theta(r_0-r)$:
\begin{equation}
\delta E_{-j} =\langle \Psi^{LLL}_{-j} |V(r)|\Psi^{LLL}_{-j} \rangle= -\frac{V_0}{l^2_B j!}\int^{r_0}_0 \left(\frac{r^{2}}{2l^2_B}\right)^{j}e^{-r^2/(2l^2_B)}\,rdr.
\label{app-dE}
\end{equation}
It equals
\begin{equation}
\delta E_{-j} \approx -\frac{V_0}{(j+1)!}\left(\frac{r^2_0}{2l^2_B}\right)^{j+1}, \qquad r^2_0 \ll 2l^2_B
\end{equation}
and
\begin{equation}
\delta E_{-j} \approx -V_0\left[1-\left(\frac{r^2_0}{2l^2_B}\right)^{j}\frac{e^{-r^2_0/(2l^2_B)}}{j!}\right], \qquad r^2_0 \gg 2l^2_Bj.
\label{graphene-correction}
\end{equation}
Note that $\delta E_{-j}$ increases (becomes less negative) with $j$ for a narrow potential well and for a wide potential well in the case where $j$ is not too large, $r^2_0 \gg 2l^2_B j$, so that the asymptotics used in Eq.~(\ref{graphene-correction}) are applicable. The limit $r_0 \to \infty$ is quite transparent in graphene as $\delta E_{-j}$ tends to $-V_0$, which simply corresponds to a shift of energy. As to the first-order correction to the wave function in the LLL approximation when mixing with higher LLs is neglected, it vanishes because the matrix element of the potential is zero, $V_{jj^{\prime}}=0$, for $j^{\prime} \ne j$ due to the angular integral. Thus, the localization of bound states for $V_0 \to 0$ is defined by the unperturbed wave function given in Eq.~(\ref{app-ab}) and does not show any enhanced localization at the potential well edge $r_0$ at large $|j|$ in agreement with the absence of singularity of the interaction term $f^{\prime}V^{\prime}/(\epsilon-V-\Delta)$  at the turning point in Eq.~(\ref{equation-generic-1}). As we showed in the main text, this is not the case for pseudospin-1 fermions.

\end{widetext}

\bibliography{Library-short}

\end{document}